\documentclass[reprint,superscriptaddress,amsmath,amssymb,aps,prl, longbibliography]{revtex4-1}

\usepackage{graphicx}
\usepackage{bm}
\usepackage{hyperref}
\usepackage{xr-hyper}
\usepackage{xcolor}
\makeatletter
\newcommand*{\addFileDependency}[1]{% argument=file name and extension
  \typeout{(#1)}
  \@addtofilelist{#1}
  \IfFileExists{#1}{}{\typeout{No file #1.}}
}
\makeatother

\newcommand*{\myexternaldocument}[1]{%
    \externaldocument{#1}%
    \addFileDependency{#1.tex}%
    \addFileDependency{#1.aux}%
}

\myexternaldocument{supp}

\begin{document}

\title{Electronic layer decoupling driven by density-wave order in La$_4$Ni$_3$O$_{10}$}

\author{Ziqiang Guan}
\thanks{these authors contributed equally}
\affiliation{Department of Physics, Harvard University, Cambridge, Massachusetts 02138, USA}

\author{Sophia F. R. TenHuisen}
\thanks{these authors contributed equally}
\affiliation{Department of Physics, Harvard University, Cambridge, Massachusetts 02138, USA}
\affiliation{John A. Paulson School of Engineering and Applied Sciences, Harvard University, Cambridge, Massachusetts 02138, USA}

\author{M. Tepie}
\affiliation{Department of Physics, Harvard University, Cambridge, Massachusetts 02138, USA}

\author{Yifeng Zhao}
\author{Ezra Day-Roberts}
\author{Harrison LaBollita}
\affiliation{Department of Physics, Arizona State University, Tempe, Arizona 85287, USA}

\author{Alexander M. Young}
\author{Xiaomeng Cui}
\affiliation{Department of Physics, Harvard University, Cambridge, Massachusetts 02138, USA}

\author{Xinglong Chen}
\affiliation{Materials Science Division, Argonne National Laboratory, Lemont, Illinois 60439, USA}

\author{Filippo Glerean}
\affiliation{Department of Physics, Harvard University, Cambridge, Massachusetts 02138, USA}
\affiliation{Condensed Matter Physics and Materials Science Department, Brookhaven National Laboratory, Upton, New York 11973, USA}

\author{Carl Audric Guia}
\affiliation{Department of Physics, Harvard University, Cambridge, Massachusetts 02138, USA}

\author{Mark P. M. Dean}
\affiliation{Condensed Matter Physics and Materials Science Department, Brookhaven National Laboratory, Upton, New York 11973, USA}

\author{Philip Kim}
\affiliation{Department of Physics, Harvard University, Cambridge, Massachusetts 02138, USA}
\affiliation{John A. Paulson School of Engineering and Applied Sciences, Harvard University, Cambridge, Massachusetts 02138, USA}

\author{J. F. Mitchell}
\affiliation{Materials Science Division, Argonne National Laboratory, Lemont, Illinois 60439, USA}

\author{Antia S. Botana}
\affiliation{Department of Physics, Arizona State University, Tempe, Arizona 85287, USA}

\author{Christopher C. Homes}
\affiliation{National Synchrotron Light Source II, Brookhaven National Laboratory, Upton, NY, USA}

\author{Matteo Mitrano}
\email{mmitrano@g.harvard.edu}
\affiliation{Department of Physics, Harvard University, Cambridge, Massachusetts 02138, USA}

\date{\today}

\begin{abstract}
We probe the density-wave transition of the trilayer nickelate La$_4$Ni$_3$O$_{10}$ with polarization-resolved infrared spectroscopy. The low-energy electrodynamics is strongly anisotropic, with metallic in-plane and insulating out-of-plane character. In the ordered phase, the anisotropy grows more than an order of magnitude as the out-of-plane conductivity is sharply suppressed. We interpret this enhancement as an effective electronic decoupling of the Ni-O layers, driven by a spin-density-wave--induced redistribution of Ni-$d_{z^2}$ occupation within the trilayers. This electronic response is accompanied by clear shifts and splittings of the out-of-plane phonons, compatible with a density-wave instability of electronic origin.
\end{abstract}

\maketitle

Bilayer La$_3$Ni$_2$O$_7$ and trilayer La$_4$Ni$_3$O$_{10}$ Ruddlesden-Popper (RP) nickelates become superconductors under pressure~\cite{sun2023signatures, zhang2024high, zhu2024superconductivity, wang2024normal, sakakibara2024theoretical, zhang2025superconductivity, li2024signature, puphal2025superconductivity, wang2025recent}. In both of these materials, superconductivity emerges from density-wave (DW) ground states at ambient pressure, seemingly emulating the phenomenology of hole-doped copper oxides~\cite{chen2024evidence, xie2024strong, khasanov2025pressure, zhang2020intertwined, khasanov2025identical, xu2025collapse}. However, unlike high-$T_c$ cuprates, whose low-energy physics is often captured by a single $d_{x^2-y^2}$ band, RP nickelates are multiorbital systems with substantial interlayer coupling. It is therefore crucial to resolve how the active orbitals contribute to the DW instabilities, mediate coupling between Ni-O planes, and influence the transport properties of these materials~\cite{wang2025recent, gim2025orbital, zhang2025spindensity}. 

\begin{figure*}
    \centering
    \includegraphics[width=\textwidth]{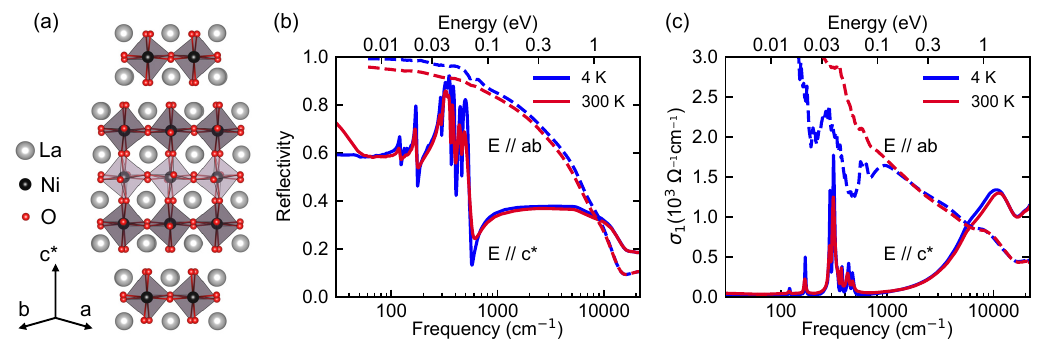}
    \caption{(a) Structure of the trilayer Ruddlesden-Popper nickelate La$_4$Ni$_3$O$_{10}$, consisting of three NiO$_3$ octahedral layers, where the inner and outer layers are colored differently to reflect their distinct local electronic environment. (b) Reflectivity and (c) real part of the optical conductivity $\sigma_1$ at base (blue) and room temperature (red). Out-of-plane ($E\parallel c^*$) and in-plane ($E\parallel ab$) optical responses are shown in solid lines and broken lines, respectively. The DW transition temperature is approximately 140~K.}
    \label{fig:F1_basic}
\end{figure*}

The trilayer RP compound La$_4$Ni$_3$O$_{10}$, shown in Fig.~\ref{fig:F1_basic}(a), is a compelling platform to address these multiorbital effects. This material has an average Ni valence of $3d^{7.33}$, and both $d_{x^2-y^2}$ and $d_{z^2}$ orbitals contribute to its low-energy physics~\cite{chen2024trilayer, yang2024effective, zhang2024prediction, labollita2024electronic, leonov2024electronic, tian2024effective}. While $d_{x^2-y^2}$ orbitals are key to the in-plane physics, $d_{z^2}$ states couple the three NiO$_3$ octahedral layers via apical-oxygen hybridization. This out-of-plane coupling also differentiates inner (light purple) and outer (dark purple) Ni-O layers due to their distinct local electronic environments~\cite{huang2024interlayer, labollita2024electronic, tian2024effective, wang2024non}. At ambient pressure, La$_4$Ni$_3$O$_{10}$ exhibits an intertwined spin- and charge-density-wave (SDW, CDW) transition near $T_{\text{DW}}\approx 140$~K. The CDW is in phase across all three layers, while the SDW resides on the two antiferromagnetically-coupled outer layers, leaving a node in the inner layer~\cite{zhang2020intertwined, labollita2024electronic, zhang2025spindensity}. Under pressure, CDW and SDW states collapse, with superconductivity appearing at T$_c\approx20-40$~K~\cite{zhu2024superconductivity, zhang2025superconductivity, khasanov2025identical}.

In this Letter, we use infrared spectroscopy to investigate how the DW phase influences the low-energy electrodynamics in trilayer nickelate La$_4$Ni$_3$O$_{10}$ bulk single crystals. Details on single-crystal sample growth, optical measurements, and computational simulations can be found in the Supplementary Material ~\cite{supp}. We observe a pronounced electronic anisotropy at far-infrared frequencies, as the in-plane conductivity is metallic, while the conductivity in the out-of-plane stacking $c^*$ direction is insulating. Across the DW transition, the out-of-plane transport is strongly suppressed, yielding an order-of-magnitude enhancement of the transport anisotropy. We attribute this effect to a DW-driven redistribution of Ni $d_{z^2}$ orbital occupation that effectively decouples the Ni-O layers. We also observe distinct phonon renormalizations and mode splitting, consistent with the presence of electron–phonon and magneto-elastic coupling. Together, these results show that the DW phase dramatically reshapes the normal-state properties of RP nickelates.

Figure~\ref{fig:F1_basic}(b) shows the polarized infrared reflectivity of La$_4$Ni$_3$O$_{10}$ for in-plane (E~$\parallel ab$) and out-of-plane (E~$\parallel c^*$) directions over a broad frequency range. Here, $c^*$ denotes the direction normal to the $ab$-plane, which is slightly tilted relative to the crystallographic $c$ axis in the monoclinic $P2_1/a$ structure. The optical response of this trilayer nickelate is strongly anisotropic. The in-plane reflectivity is metallic, exceeding 0.9 below 500~cm$^{-1}$ and exhibiting a clear plasma edge around 10,000~cm$^{-1}$. In contrast, the out-of-plane reflectivity is nearly insulating, with only a minor conductivity contribution (likely from hopping conduction) and several infrared-active phonons. This qualitative dichotomy mirrors layered cuprates, where metallic CuO$_2$ planes coexist with a nearly insulating interlayer response~\cite{basov2005electrodynamics}. Fig.~\ref{fig:F1_basic}(c) shows the optical conductivity from Kramers-Kronig transformations~\cite{dressel2002electrodynamics, tanner2015use} (see Supplementary Material Sec.~S1). The optical absorption reveals that the large electronic anisotropy surprisingly reverses around 5000 cm$^{-1}$. At low frequencies the in-plane conductivity dominates, consistent with quasi-two-dimensional metallicity, whereas at higher energies the out-of-plane conductivity exceeds the in-plane component. This crossover signals a clear energy-scale separation of orbital excitations. The high-energy response is set by strong out-of-plane coupling mediated by Ni $d_{z^2}$ orbitals, while the low-energy electrodynamics is dominated by in-plane Ni $d_{x^2-y^2}$, with a substantially reduced $d_{z^2}$ contribution~\cite{labollita2024electronic}.

The observed frequency-dependent reversal in the optical anisotropy is unusual among layered correlated materials. In cuprate superconductors, the infrared optical response is dominated by in-plane $d_{x^2-y^2}$ orbitals, while the out-of-plane conductivity is suppressed by roughly an order of magnitude across the entire infrared range without signs of anisotropy reversals~\cite{basov2005electrodynamics, uchida1996c, homes1995optical, cooper1993optical, schutzmann1994c}. In iron-based superconductors, despite their multiorbital and more three-dimensional electronic structure, the in-plane conductivity still exceeds the out-of-plane response, with anisotropy ratios typically ranging between 1 and 10 and, again, no reported reversal~\cite{chen2010measurement, cheng2011three, moon2011incoherent, moon2013interlayer}. The anisotropy reversal of our trilayer nickelate thus reflects a marked multiorbital nature, where distinct orbitals dominate the electronic properties at different energy scales, akin to earlier reports on La$_{2-x}$Sr$_x$NiO$_4$ and theoretical predictions for La$_3$Ni$_2$O$_7$~\cite{shinomori2002orientation, geisler2024optical}.

Below the DW transition, the in-plane optical conductivity remains metallic, but shows a partial mid-infrared depletion of spectral weight associated with opening of a DW gap [Fig.~\ref{fig:F1_basic}(c)]. We extract the DW gap by analyzing the differential optical conductivity (see Supplementary Sec.~S3). At base temperature, we obtain an energy gap $2\Delta_{\mathrm{DW}}=112$~meV, corresponding to $2\Delta_{\mathrm{DW}}/k_B T_{\mathrm{DW}}\approx 9.3$. This ratio is well above the weak-coupling value of 3.52, yet comparable to that of La$_3$Ni$_2$O$_7$ ~\cite{liu2024electronic}, and consistent with earlier estimates for La$_4$Ni$_3$O$_{10}$~\cite{xu2025origin, li2025direct, gim2025orbital, suthar2025multiorbital}. A full temperature dependence, shown in Supplementary Fig.~S1, verifies that this change occurs continuously. The gap $\Delta(T)$ follows a mean-field behavior, in line with prior X-ray diffraction and neutron scattering~\cite{zhang2020intertwined}, and points to a second-order transition, likely driven by an SDW instability~\cite{norman2025landau}.

\begin{figure}
    \centering
    \includegraphics[width=0.51\textwidth]{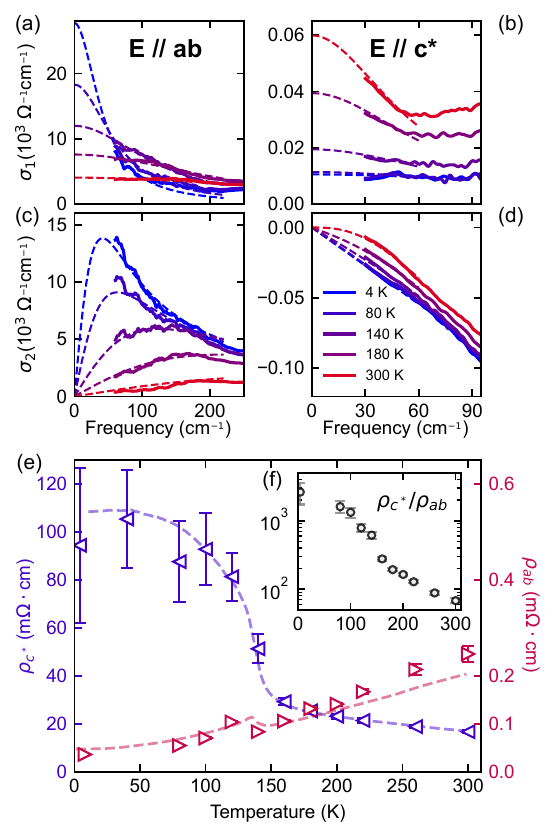}
\caption{ (a)-(d) Temperature-dependent real and imaginary optical conductivity of La$_4$Ni$_3$O$_{10}$ in- ($E\parallel ab$) and out-of-plane ($E\parallel c^*$), denoted as solid lines. Drude$+\varepsilon_\infty$ model fits are shown as dashed lines. (e) Temperature dependence of the extrapolated DC resistivity obtained from the Drude fits. Left- and right-pointing triangles with error bars represent the out-of-plane ($\rho_{c^*}$) and in-plane ($\rho_{ab}$) resistivities, respectively. Broken lines denote DC resistivity obtained from transport measurements. $\rho_{ab}$ is adapted from Ref.~\cite{zhang2020high} (see Supplementary Material Sec.~S6 for further details). (f) Inset: Temperature-dependent resistivity anisotropy ratio $\rho_{c^*}/\rho_{ab}$.}
    \label{fig:F2_anisotropy_Tdep}
\end{figure}

Figures~\ref{fig:F2_anisotropy_Tdep}(a)-(d) show the main result of our work, where we observe a dramatic effect of the density wave on the temperature dependence of the low-frequency in- and out-of-plane optical conductivity. At frequencies below the phonon and interband contributions, we simultaneously fit the real and imaginary optical conductivity with a simplified Drude$+\varepsilon_\infty$ model, and obtain extrapolated DC resistivities $\rho_{ab}$ and $\rho_{c^*}$ from $\rho_{\mathrm{dc}}=\lim_{\omega\rightarrow 0}1/\sigma_1(\omega)$ at each temperature [Fig.~\ref{fig:F2_anisotropy_Tdep}(e)]. While $\rho_{ab}$ remains metallic across the entire temperature range, showing only a weak kink at $T_{\mathrm{DW}}$, $\rho_{c^*}$ rises abruptly by nearly a factor of five upon cooling through $T_{\mathrm{DW}}$. As a result, the anisotropy ratio $\rho_{c^*}/\rho_{ab}$ [Fig.~\ref{fig:F2_anisotropy_Tdep}(f), inset] grows by over an order of magnitude, from $\rho_{c^*}/\rho_{ab}\simeq 70$ at room temperature to $\rho_{c^*}/\rho_{ab}\simeq 2600$ at 4~K, with the sharpest enhancement occurring at the DW transition. We also note that these optically extrapolated DC resistivities agree with separately measured DC transport measurement results (shown as broken lines in Fig. 2(e)). These results show that the DW transition drives a strongly anisotropic reorganization of charge dynamics, with out-of-plane transport suppressed far more than that in-plane. The large and strongly temperature-dependent anisotropy is consistent with an effective dimensional crossover, meaning that the electronic properties transition from a moderately three-dimensional character at high-temperature to a highly two-dimensional character in the low-temperature DW phase.

\begin{figure}
    \centering
    \includegraphics[width=0.48\textwidth]{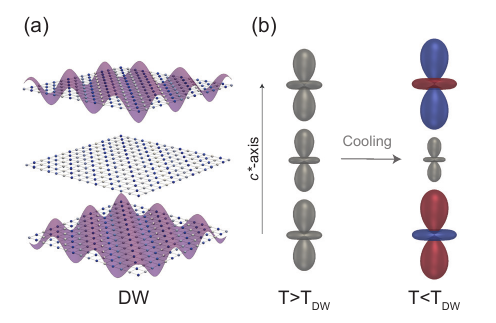}
    \caption{(a) Real-space depiction of the DW modulation. Below $T_{\mathrm{DW}}$, an antiphase SDW develops on the outer layers, leaving the inner layer as a magnetic node, while the intertwined CDW is in-phase across the trilayer. The SDW is indicated by the purple wave, and the CDW by the blue/white variation of the lattice color. 
    (b) Proposed qualitative schematic of DW-driven redistribution of Ni $d_{z^2}$ orbital occupation within a trilayer. Above $T_{\mathrm{DW}}$ the trilayer is nonmagnetic and the $d_{z^2}$ occupation is comparable in inner and outer Ni-O layers despite their inequivalent local environments. Below $T_{\mathrm{DW}}$, the SDW residing on the outer layers (with a magnetic node on the inner layer) is accompanied by an enhanced $d_{z^2}$ weight on the outer layers and a reduced weight on the inner layer, thereby suppressing interlayer transport. The inverted colors on the outer layer $d_{z^2}$ orbitals illustrate the phase of the SDW modulation.  }
    \label{fig:F3_conceptual}
\end{figure}

We argue that the enhanced electronic anisotropy follows from an effective decoupling of the layers within each trilayer subunit (Fig.~\ref{fig:F3_conceptual}). In trilayer RP nickelates, strong interlayer hybridization of the Ni $d_{z^2}$ orbitals produces three molecular subbands (bonding, nonbonding, antibonding) quantum-confined by La–O spacer layers between successive Ni–O trilayers. The bonding and antibonding states carry $d_{z^2}$ weight on all three layers, whereas the odd-symmetry nonbonding band has a node on the inner Ni layer. As a result, layer-differentiated filling of the $d_{z^2}$ orbitals emerges naturally in this system, independent of temperature \cite{pardo2010quantum, jung2022electronic, labollita2024electronic, leonov2024electronic, wang2024non}. Upon cooling into the DW state, La$_4$Ni$_3$O$_{10}$ develops an intertwined SDW–CDW state in which the SDW resides on the two outer Ni–O layers, leaving a node on the nonmagnetic inner layer~\cite{zhang2020intertwined} [Fig.~\ref{fig:F3_conceptual}(a)]. Concomitantly, the DW redistributes orbital weight across the trilayer and promotes the nonbonding component, thus increasing $d_{z^2}$ occupation on the outer layers while lowering it on the inner layer~\cite{labollita2024electronic, zhang2025spin} [Fig.~\ref{fig:F3_conceptual}(b)]. This redistribution reduces the effective interlayer overlap of the $d_{z^2}$ orbitals, and suppresses out-of-plane charge transport.

\begin{figure*}[htbp]
    \centering
    \includegraphics[width=0.98\textwidth]{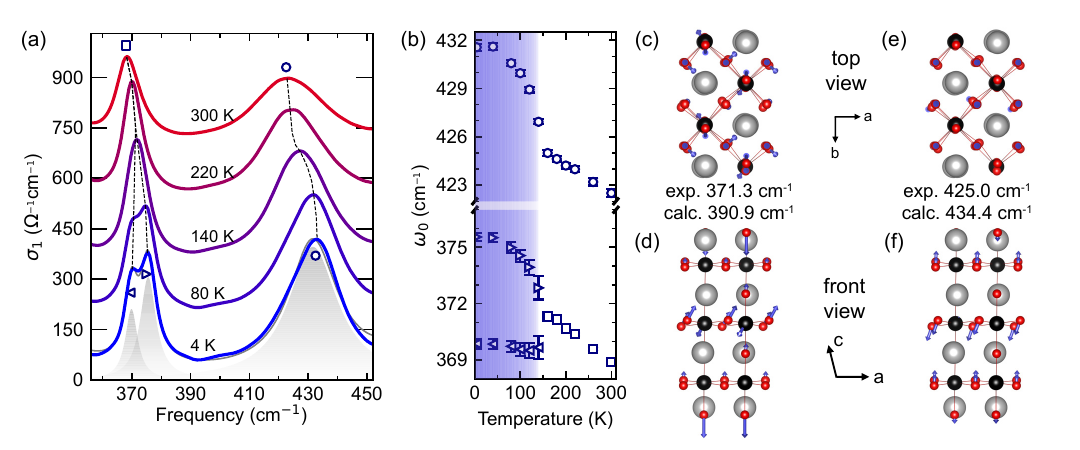}
    \caption{(a) Temperature evolution of the out-of-plane vibrational response. Spectra are vertically offset by 150~$\Omega^{-1}$cm$^{-1}$ for clarity. Symbols mark phonon frequencies and dotted lines are guides to the eye. Drude-Lorentz fits to the 4 K spectrum with individual contributions are shown as grey lines and shaded areas, respectively. (b) Temperature-dependent phonon frequencies for selected modes; triangles denote split modes in the DW phase (blue shading). (c)-(f) Calculated atomic displacements for infrared-active modes matching the 160 K data in panel (b); only large-amplitude displacements are shown for clarity. Owing to the monoclinic angle $\beta$, the crystallographic $c$ axis is tilted by $\sim 11^{\circ}$ relative to the experimental out-of-plane $c^*$ direction. (c)-(d) and (e)-(f) are views of the same mode from two different directions.}
    \label{fig:F4_phonon_anomaly}
\end{figure*}

This electronic layer decoupling is supported by the kinetic-energy renormalization $K_{\mathrm{exp}}/K_{\mathrm{DFT}}$ (Supplementary Sec.~\S4), where $K_{\mathrm{exp}}$ is obtained from the Drude plasma frequency and $K_{\mathrm{DFT}}$ from uncorrelated DFT. At 160~K (above $T_{\mathrm{DW}}$) we find $K^{ab}_{\mathrm{exp}}/K^{ab}_{\mathrm{DFT}}=0.237$, consistent with a moderately correlated metal~\cite{xu2025origin, liu2025evolution}, while the $c^*$-axis renormalization is far stronger, $K^{c^*}_{\mathrm{exp}}/K^{c^*}_{\mathrm{DFT}}=0.032$. The resulting anisotropy implies that $c^*$-axis charge dynamics, set by $d_{z^2}$-mediated interlayer hopping, is already strongly correlated at high temperature and thus especially susceptible to the DW transition, consistent with the observed layer decoupling. Changes in quasiparticle coherence~\cite{su2024strongly} and/or a possible partial gap opening on $d_{z^2}$-derived states~\cite{li2017fermiology} may further suppress the low-energy $c^*$-axis response in the DW state.

The pronounced enhancement of the low energy electronic anisotropy highlights the distinctive place of RP nickelates within the landscape of unconventional superconductors. Cuprates typically display very large, strongly temperature-dependent electronic anisotropy, sometimes accompanied by signatures of a dimensional crossover toward a highly two-dimensional low-energy state~\cite{nakamura1993anisotropic,komiya2002c, martin1988temperature, watanabe1997anisotropic, zavaritsky2005normal, takenaka1994interplane, hussey1997anisotropic, lavrov1998normal}. However, their physics is effectively described by a single $d_{x^2-y^2}$ orbital. By contrast, iron-based superconductors, where multiple orbitals contribute comparably to both in-plane and out-of-plane transport, exhibit only moderate anisotropy with weak temperature dependence~\cite{wang2009anisotropy, chen2008transport, tanatar2009resistivity, nakajima2018comprehensive, moon2011incoherent, wang2011superconductivity, lei2011anisotropy, vedeneev2013temperature, song2010synthesis}. Our data show that La$_4$Ni$_3$O$_{10}$ realizes an intermediate regime where several orbitals remain active near $E_F$ and interlayer coupling is appreciable. Yet once DW order sets in, it sharply amplifies the transport anisotropy by effectively decoupling the Ni-O layers within each trilayer subunit, enhancing the two-dimensional character of the low-energy charge dynamics~\cite{chen2025nickelate, puphal2025superconductivity}. 

Beyond reshaping the low-energy electrodynamics, the DW transition also imprints clear signatures in the $c^*$-axis vibrational spectrum. Analysis of the infrared-active phonons (see Supplementary Sec.~S9) shows that the strongest effects occur between 350 and 450~cm$^{-1}$ [Figs.~\ref{fig:F4_phonon_anomaly}(a)-(b)]. Upon cooling into the DW state, the 370~cm$^{-1}$ mode splits into two branches, while the 430~cm$^{-1}$ mode strongly hardens without splitting [Fig.~\ref{fig:F4_phonon_anomaly}(b)]. Additionally, weaker anomalies are discussed in the Supplementary Material Sec.~S9, but in all cases the low-temperature evolution departs significantly from the smooth anharmonic trends above $T_{\mathrm{DW}}$. We compared the room-temperature results to first-principles calculations for the monoclinic $P2_1/a$ crystal symmetry. Based on this analysis, we associate the phonons at 371 and 425 cm$^{-1}$ to the calculated eigenvectors in Figs.~\ref{fig:F4_phonon_anomaly}(c)-(f). These are dominated by Ni-O bond distortions with mixed bending and stretching, implying that the DW primarily perturbs lattice distortions coupled to charge modulation in the Ni-O layers.

These vibrational anomalies could originate from (i) a structural transition, (ii) zone folding by the DW superlattice, (iii) electron-phonon coupling, or (iv) magnetoelastic coupling. Across the DW transition, the crystal structural symmetry remains unchanged and the reported lattice-parameter anomalies are well below the 0.1\% level~\cite{zhang2020high, kumar2020physical, rout2020structural, li2025crystal}, yet we observe $\sim$1\% frequency shifts and splittings only for certain phonons in a frequency window of 300-500~cm$^{-1}$, far exceeding expectations from such small structural changes. This mismatch, together with the absence of structural symmetry lowering, disfavors an explanation based on a simple lattice instability. Zone folding by the DW superlattice, which back-folds finite-momentum phonons to the zone center~\cite{deswal2025dynamics}, is likewise inconsistent with our data. The DW in La$_4$Ni$_3$O$_{10}$ is incommensurate, so any folded spectral weight would be broadly distributed and likely too weak to yield sharp new modes. Even approximating the DW by its closest five-unit-cell periodicity would generically activate many additional modes across the spectrum. Instead, we observe clear DW anomalies and only in a small subset of Ni-O vibrations. Electron-phonon coupling in the DW state is a more compelling mechanism.
The CDW is out of phase between neighboring trilayers~\cite{zhang2020intertwined}, effectively doubling the electronic periodicity along $c^*$ and introducing a mode degeneracy that can be lifted by DW-driven electronic renormalization. The unidirectional CDW in the Ni–O layers creates an anisotropic charge distribution~\cite{zhang2020intertwined}, which can alter the elastic constants along orthogonal Ni–O directions and split selected Ni–O modes. Magnetoelastic coupling may also contribute by renormalizing phonon energies and lifetimes and, in some cases, producing mode splitting~\cite{jana2025strong}. Because the CDW in La$_4$Ni$_3$O$_{10}$ is intertwined with SDW order~\cite{li2025direct, norman2025landau}, and the phonon anomalies occur at energies comparable to magnetic excitations (up to $\sim$50~meV)~\cite{chen2024electronic}, coupling between lattice vibrations and collective spin excitations is a realistic possibility. 

In the aggregate, our observations establish La$_4$Ni$_3$O$_{10}$ as a strongly anisotropic, multiorbital metal. Upon entering the DW state, its low-energy anisotropy increases by over an order of magnitude, consistent with an effective decoupling of the Ni-O layers within each trilayer subunit. The simplest reading of the optical data is that this process is electronically driven, arising from a DW-induced redistribution of Ni $d_{z^2}$ occupation that quenches interlayer charge dynamics without requiring a symmetry-lowering structural transition. 
While the Ni $d_{x^2-y^2}$ and $d_{z^2}$ orbitals are energetically distinct, with $d_{x^2-y^2}$ states dominating the in-plane metallic state, we find that $d_{z^2}$ orbitals govern both the interband transition and hopping conduction along the out-of-plane direction.
Overall, the unique interplay between the two orbitals in layered nickelates defines the low-energy electrodynamics and yields an emergent quasi-two-dimensional response promoted by DW order.
The significant impact of the DW state on the lattice dynamics, in the absence of correspondingly large structural changes, further supports a picture in which the DW state is primarily electronic in origin. Our results imply that any microscopic discussion of the superconducting instability in this family should explicitly account for multiorbital effects and changes of the coupling among the layers.

\textit{Note added}---After completing this work, we became aware of a recent optical study of trilayer La$_4$Ni$_3$O$_{10}$~\cite{liu2025highly}. While the overall phenomenology is consistent, our work focuses on the effect of the DW transition on the in- and out-of plane electronic response below 100-150 cm$^{-1}$ and the vibrational properties, thereby providing complementary information on the low-energy physics of this material.

% --- Acknowledgments ---
\begin{acknowledgments}
We thank D. Nicoletti, J. Li, S. Priya for insightful discussions. Work by Z.G., S.F.R.T., M.T., F.G., C.C.H., M.P.M.D., and M.M. was supported by the U.S. Department of Energy (DOE), Division of Materials Science, under Contract No. DE-SC0012704. Part of the experimental work was performed in the Infrared Lab at the National Synchrotron Light Source II, a DOE Office of Science User Facility operated for the DOE Office of Science by Brookhaven National Laboratory. S.F.R.T. acknowledges additional support from the DOE, Office of Science, Office of Workforce Development for Teachers and Scientists, Office of Science Graduate Student Research (SCGSR) program. The SCGSR program is administered by the Oak Ridge Institute for Science and Education for the DOE under Contract No. DE-SC0014664. A.S.B and Y.Z. acknowledge support from NSF Grant No. DMR-2323971 and the ASU research computing center for HPC resources. Work by X.C. and J.F.M. in the Materials Science Division at Argonne National Laboratory (crystal growth) was supported by the U.S. Department of Energy Office of Science, Basic Energy Sciences, Materials Science and Engineering Division. Work by X.C. and P.K. was supported by AFOSR (FA9550-25-1-0019). A.M.Y. was supported by NSF (DMR-2105048). 
\end{acknowledgments}

% --- Data/Code availability (optional for PRL, often moved to Supplemental) ---
% \paragraph*{Data Availability}\;Data and code are available at: URL/DOI.

% --- Bibliography ---
% Use a .bib file and apsrev4-2 style. Titles are optional in PRL.
\bibliography{ref} % refs.bib

% --- Supplemental Material (if needed) ---
% Create a separate file or use the \widetext environment sparingly.
% \onecolumngrid
% \section*{Supplemental Material}
% Additional methods, extended data, and derivations.
% \twocolumngrid

\end{document}

\title{Supplemental Material for: Electronic layer decoupling driven by density-wave order in La$_4$Ni$_3$O$_{10}$}

\author{Ziqiang Guan}
\affiliation{Department of Physics, Harvard University, Cambridge, Massachusetts 02138, USA}

\author{Sophia F. R. TenHuisen}
\affiliation{Department of Physics, Harvard University, Cambridge, Massachusetts 02138, USA}
\affiliation{John A. Paulson School of Engineering and Applied Sciences, Harvard University, Cambridge, Massachusetts 02138, USA}

\author{M. Tepie}
\affiliation{Department of Physics, Harvard University, Cambridge, Massachusetts 02138, USA}

\author{Yifeng Zhao}
\author{Ezra Day-Roberts}
\author{Harrison LaBollita}
\affiliation{Department of Physics, Arizona State University, Tempe, Arizona 85287, USA}

\author{Alexander M. Young}
\author{Xiaomeng Cui}
\affiliation{Department of Physics, Harvard University, Cambridge, Massachusetts 02138, USA}

\author{Xinglong Chen}
\affiliation{Materials Science Division, Argonne National Laboratory, Lemont, Illinois 60439, USA}

\author{Filippo Glerean}
\affiliation{Department of Physics, Harvard University, Cambridge, Massachusetts 02138, USA}
\affiliation{Condensed Matter Physics and Materials Science Department, Brookhaven National Laboratory, Upton, New York 11973, USA}

\author{Carl Audric Guia}
\affiliation{Department of Physics, Harvard University, Cambridge, Massachusetts 02138, USA}

\author{Mark P. M. Dean}
\affiliation{Condensed Matter Physics and Materials Science Department, Brookhaven National Laboratory, Upton, New York 11973, USA}

\author{Philip Kim}
\affiliation{Department of Physics, Harvard University, Cambridge, Massachusetts 02138, USA}
\affiliation{John A. Paulson School of Engineering and Applied Sciences, Harvard University, Cambridge, Massachusetts 02138, USA}

\author{J. F. Mitchell}
\affiliation{Materials Science Division, Argonne National Laboratory, Lemont, Illinois 60439, USA}

\author{Antia S. Botana}
\affiliation{Department of Physics, Arizona State University, Tempe, Arizona 85287, USA}

\author{Christopher C. Homes}
\affiliation{National Synchrotron Light Source II, Brookhaven National Laboratory, Upton, NY, USA}

\author{Matteo Mitrano}
\affiliation{Department of Physics, Harvard University, Cambridge, Massachusetts 02138, USA}

% Optional corresponding author
% \email{corresponding.author@institute.edu}

\date{\today}

\maketitle

\section{Sample Preparation, Optical Experimental Methods, and Full Temperature-dependent Dataset} \label{sec:expmethod}
High-quality single crystals of La$_4$Ni$_3$O$_{10}$ were grown using a vertical optical-image floating zone furnace operating at high O$_2$ pressure according to the procedures described in Ref.~\cite{zhang2020high}. The quality and orientation of the single crystals were confirmed by x-ray Laue diffraction.
Two large single crystals (approximately $3\times2\times2$~mm$^3$) of La$_4$Ni$_3$O$_{10}$ were polished to expose an $ab$-plane and an $ac^*$-plane, respectively. The near-normal-incidence reflectivity $R(\omega)$ at ambient pressure was measured in the frequency range of 60-22,000~cm$^{-1}$ using a Bruker Vertex 80v Fourier-transform infrared spectrometer (FTIR) and referenced with in situ gold evaporation \cite{homes1993technique}. 

We used unpolarized light to measure the $ab$-plane reflectivity on an $ab$-plane–polished sample. Polishing introduced a small misorientation of the crystal surface ($\sim$2$^\circ$), as verified by subsequent x-ray diffraction on the same sample. This misorientation slightly mixes $c^*$-axis signatures into the $ab$-plane FTIR measurement with unpolarized light, which in practice appears as a small reduction of the overall reflectivity and weak phonon-like peaks at the same frequencies as in the $c^*$-axis spectrum~\cite{homes2007stripe}. We correct for this mixing by subtracting the $c^*$-axis reflectivity contribution and renormalizing the result. This correction does not affect the conclusions of this article. Owing to the high $ab$-plane conductivity and reflectivity, we use a Hagen-Rubens extrapolation below the lowest measured frequency.

We used out-of-plane–polarized light to measure the $c^*$-axis reflectivity on an $ac^*$-plane–polished sample. For this configuration, we also performed polarization-resolved terahertz time-domain spectroscopy (THz-TDS) to extend the low-frequency data down to 30~cm$^{-1}$. We employed an echelon-based single-shot THz-TDS system \cite{padma2025symmetry, gao2022high} pumped by 800~nm, 35~fs pulses from a 1~kHz Ti:sapphire amplifier. THz pulses were generated in a 0.5-mm-thick $\braket{110}$ ZnTe crystal and electro-optic sampling of the light reflected from the sample was carried out focusing on a 0.2-mm-thick $\langle110\rangle$ ZnTe sensor optically contacted to a 1-mm-thick $\langle100\rangle$ ZnTe substrate. The THz reflectivity was obtained by normalizing the THz spectrum of the sample to that of a gold reference measured under identical conditions. The entire THz-TDS apparatus operated under vacuum. Because the $c^*$-axis conductivity and reflectivity are lower and the low-frequency region contains many Lorentz modes, we fit the low-frequency reflectivity with a Drude model including a background dielectric permittivity (a simple $\varepsilon_\infty+$Drude approximation) and use it to extrapolate below the lowest measured frequency~\cite{tanner2015use}.

The complex optical conductivity was determined via a Kramers-Kronig analysis of the measured reflectivity. For both the $ab$-plane and $c^*$-axis reflectivity, we use an x-ray atomic scattering factor for the high-frequency extrapolation\cite{tanner2015use}. We also tested different high-frequency extrapolation methods, such as assuming a constant reflectivity up to $10^6$ cm$^{-1}$ ($\sim$ 124\;eV) followed by a free-electron ($\omega^{-4}$) response. The complex optical conductivity calculated based on different methods shows no qualitative difference in the measured frequency range.

\begin{figure}
    \centering
    \includegraphics[width=0.98\textwidth]{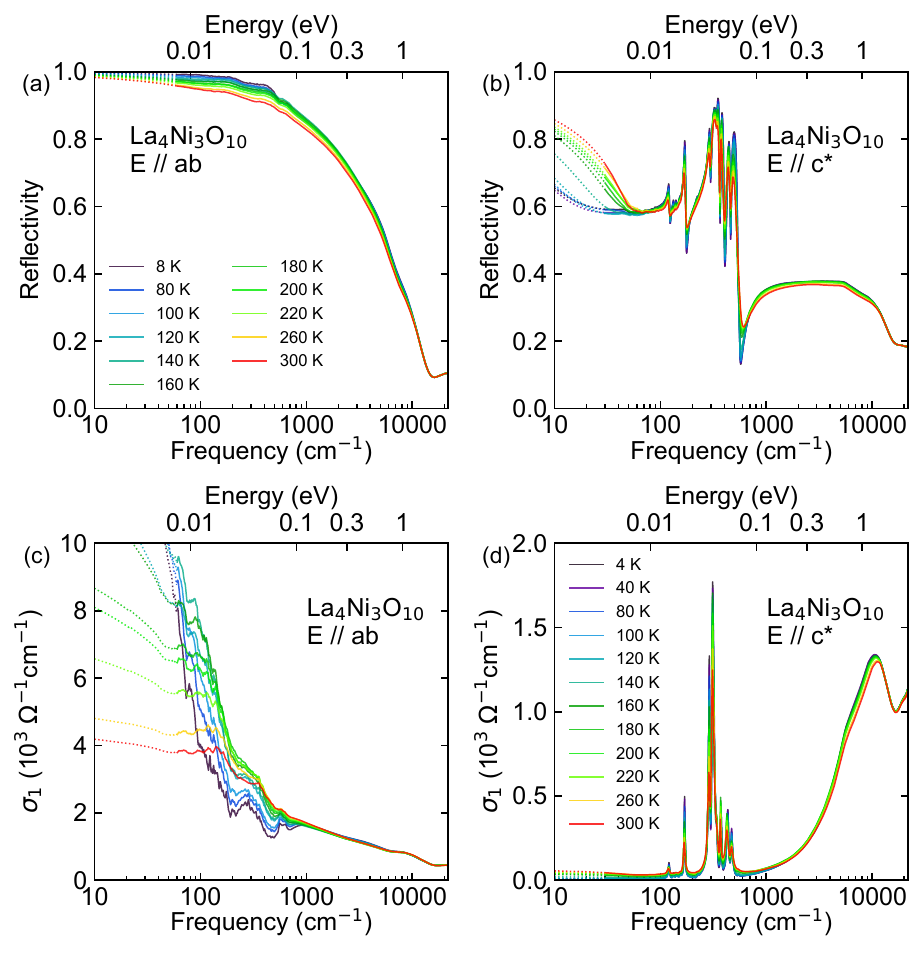}
    \caption{Reflectivity (a)(b) and real part of optical conductivity (c)(d) over a broad frequency at all measured temperatures. Out-of-plane ($E\parallel c^*$) and in-plane ($E\parallel ab$) are shown on the left and right, respectively. Dotted lines indicate the low-frequency extrapolations as described in experimental method Sec. \ref{sec:expmethod}.}
    \label{fig:S1_allT_refl_cond}
\end{figure}

Fig. \ref{fig:S1_allT_refl_cond} shows the full temperature-dependent reflectivity and conductivity data set of both in-plane and out-of-plane polarization. The $ab$-plane response shows robust metallic behavior as the reflectivity approaches unity in the zero-frequency limit and keeps rising with decreasing temperature in the far-infrared range ($<$400\,cm$^{-1}$). Suppression in the mid-infrared range (400-1000\,cm$^{-1}$), together with the spectral weight transfer to higher frequency range, indicate the onset of a density wave energy gap below the transition temperature of 140~K. For the out-of-plane response, the low frequency ($<100$\,cm$^{-1}$) reflectivity shows an upturn for temperatures above the DW transition, suggesting a finite Drude component. However, the Drude component is suppressed at low temperatures and remains relatively small compared to the Lorentzian phonon modes or the interband transitions as seen in the conductivity.

\section{DFT electronic structure and optical properties}\label{sec:dft}

The electronic structure and optical properties calculations were performed using all-electron, full-potential DFT code WIEN2K \cite{wien2k}. We used the Perdew-Burke-Ernzerhof version of the generalized gradient approximation as the exchange-correlation functional \cite{gga_pbe}. All calculations were performed using the experimental structure with $P2_{1}/a$ symmetry at ambient pressure \cite{zhang2020high} in the nonmagnetic state. A 5×13×13 k-point grid was used for the Brillouin sampling in self-consistent calculations, while a denser k-point grid of 11×29×30 was used to accurately describe the optical properties. An $RK_{\rm max}$ = 7 and muffin-tin radii of 2.38, 1.91, and 1.70 a.u. were used for La, Ni, O, respectively. The broadening factor (scattering rate) used to compute the interband contribution to the optical conductivity is set to be 0.1 eV. For the intraband contribution, the plasma frequency is directly obtained from DFT calculations, while the scattering rate (Drude width) is taken from fitted parameters in the experimental optical conductivity $\sigma$($\omega$) at 160 K. The in-plane optical conductivity is obtained by averaging the two inequivalent in-plane components.

\begin{figure}
    \centering
    \includegraphics[width=0.7\textwidth]{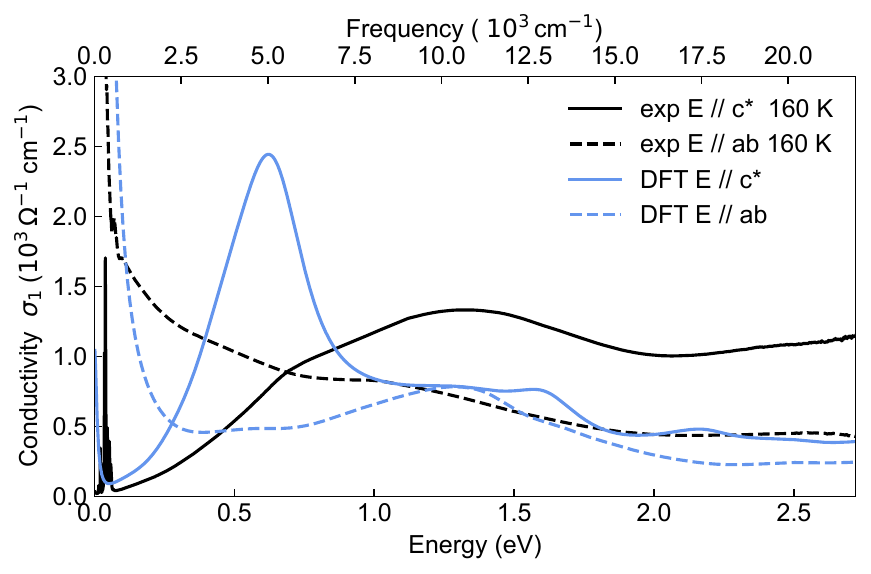}
    \caption{Comparison of optical conductivity measured experimentally (exp) and calculated with DFT. }
    \label{fig:S2_DFT_compare}
\end{figure}

Figure. \ref{fig:S2_DFT_compare} presents the comparison between experimentally measured and DFT calculated real part of the optical conductivity. The DFT calculation shown here did not include a Hubbard $U$. Similarly to the experimental result, the DFT calculation also reveals a strong anisotropy reversal as a function of frequency. The DFT calculated out-of-plane optical conductivity exceeds the in-plane components substantially around 0.6\,eV. However, the anisotropy crossover appears near 0.3\,eV, much lower than the experimentally observed crossover energy. This discrepancy is to some extent expected as general DFT calculations usually underestimate the interband transition energies \cite{seidl1996generalized}. %In La$_4$Ni$_3$O$_{10}$, the band gap is likely underestimated by standard DFT due to self-interaction errors and the missing derivative discontinuity in the exchange–correlation potential\cite{cohen2008insights}. 
Electronic-correlation-induced mass-renormalization could also contribute to the underestimation of the interband transition energy, as suggested by previous ARPES results\cite{li2017fermiology}. Further, the calculations were performed in the nonmagnetic state only and not in the DW state (where the Ni-O planes are decoupled via the nonmagnetic mirror layer). 

\begin{figure}
    \centering
    \includegraphics[width=0.85\textwidth]{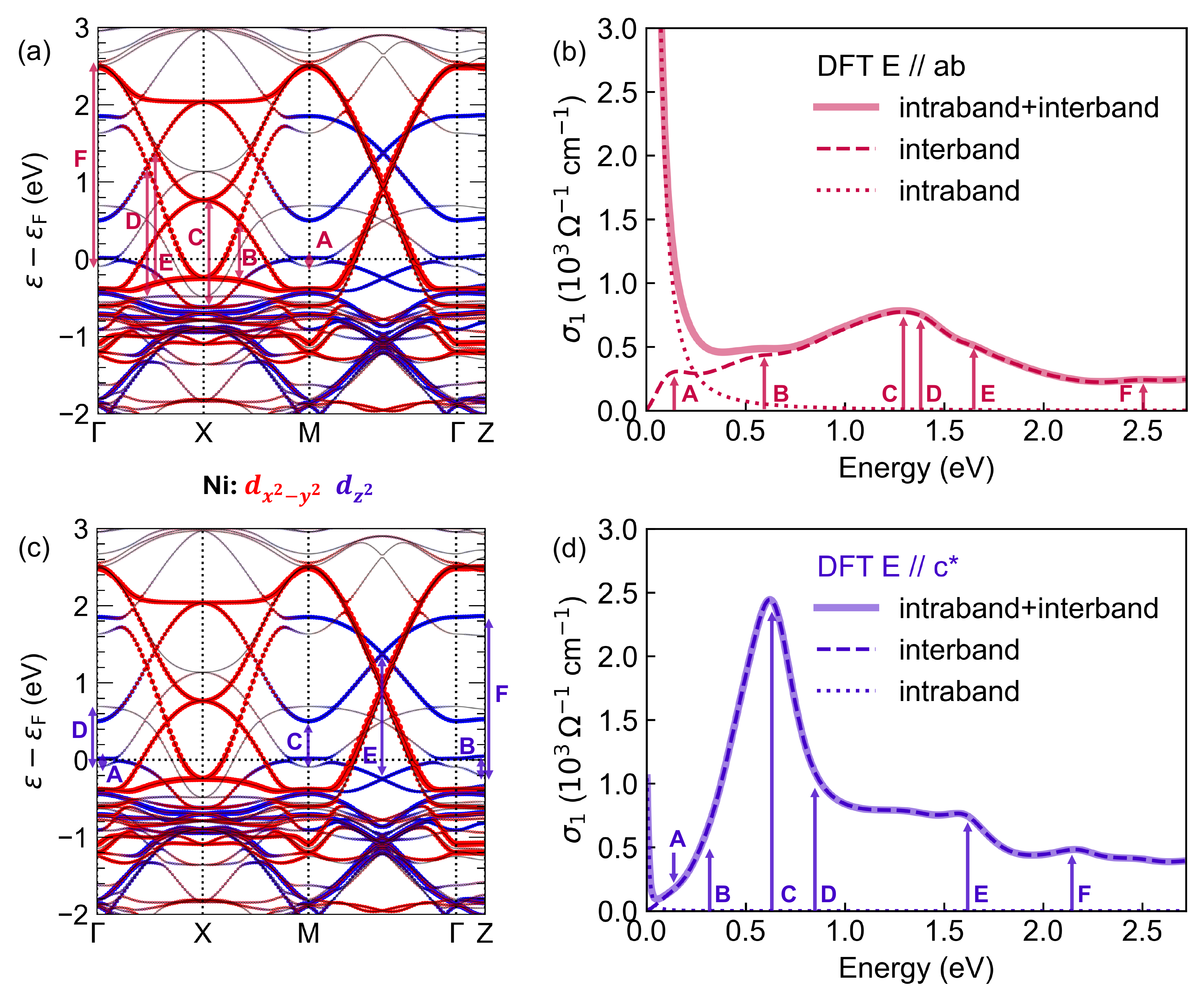}
    \caption{DFT calculated band structure and optical conductivity of La$_4$Ni$_3$O$_{10}$. Panels (a) and (c) display identical band structures, colored to highlight the dominant orbital character: Ni $d_{x^2-y^2}$ components in red and $d_{z^2}$ components in blue. These bands give rise to the distinct optical conductivities along the in-plane and out-of-plane directions, shown in (b) and (d), respectively. The total optical conductivity, along with its interband and intraband components, is presented for both polarizations. The arrows in (b) and (d) indicate the principal optical transitions, corresponding to the interband processes labeled in the band structures. The Fermi level is set to be 0 eV in panels (a) and (c).}
    \label{fig:S3_band_DFT_indentification}
\end{figure}

Next, we analyze the orbital contribution to the anisotropic optical conductivity. Figure. \ref{fig:S3_band_DFT_indentification} shows the DFT calculated band structure with Hubbard $U=0$. Near the Fermi level, Ni $d_{x^2-y^2}$ and $d_{z^2}$ bands dominates the optical transitions. By comparing the band structure with the optical conductivity, we identify that the main contributions to the in-plane interband conductivity arise from transitions between bands with dominant $d_{x^2-y^2}$ character, while the out-of-plane interband conductivity is mainly associated with transitions between bands of dominant $d_{z^2}$ character. This orbital dependence of the optical conductivity can be understood from the spatial orientation of the two $3d$ orbitals of the Ni atom. The $d_{x^2-y^2}$ orbitals lie in the Ni-O plane and mainly contribute to electronic coupling within the plane, while the $d_{z^2}$ orbitals extend along the $c^*$-axis and mainly contribute to the charge transport between layers. The difference in the orbital orientations naturally leads to the strong polarization dependence in the optical response.

\section{Density Wave Energy Gap Extraction} \label{sec:DWenergygap}
The DW optical gap appears in the in-plane conductivity as a transfer of spectral weight from low to higher frequencies. We extract the gap by analyzing the conductivity difference spectrum, $\Delta\sigma_1(\omega)=\sigma_1(\omega, T<T_{\text{DW}})-\sigma_1(\omega, T_{\text{DW}})$. We take the spectrum just above the DW transition (140~K) as the normal-state reference to minimize unrelated temperature-dependent effects.

\begin{figure}
    \centering
    \includegraphics[width=0.98\textwidth]{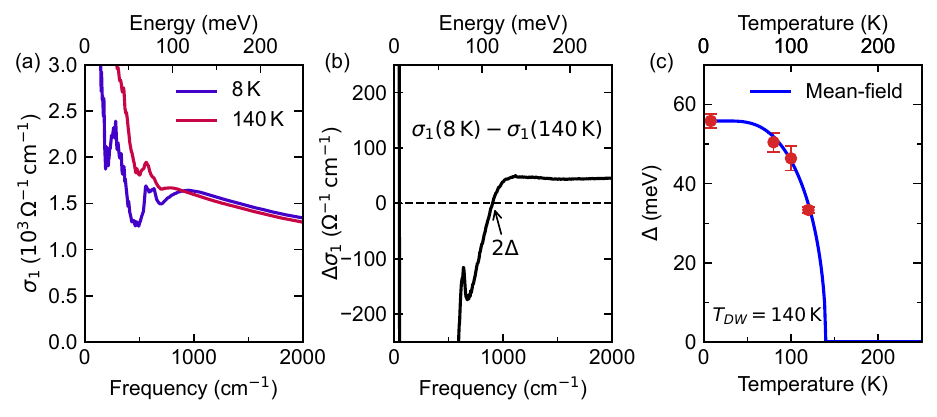}
    \caption{Density wave energy gap and mean-field behavior. (a) and (b) display the in-plane optical conductivities and their difference at the lowest measured temperature 8\,K, and just above the DW transition at 140\,K. The opening of DW energy gap $2\Delta$ is illustrated by the spectral weight transfer and crossing close to 1000\,cm$^{-1}$. The evolution of energy gap $\Delta$ with temperature is shown in panel (c). The solid blue line denotes the mean-field behavior. }
    \label{fig:S4_energy_gap_MF}
\end{figure}

Fig.~\ref{fig:S4_energy_gap_MF}(a) compares the in-plane optical conductivities at 8~K and 140~K. The spectra show a suppression in the mid-infrared and an enhancement at higher frequencies, crossing near 1000~cm$^{-1}$. We attribute this spectral-weight transfer to the opening of the DW gap, since a simple narrowing of the Drude peak cannot account for the high-frequency enhancement. Fig.~\ref{fig:S4_energy_gap_MF}(b) shows the 8~K difference conductivity $\Delta\sigma_1(\omega)$; the zero-crossing point (black arrow) defines the optical gap, $2\Delta=112$~meV. The resulting DW gap, $\Delta=56$~meV, agrees with a previous FTIR report on the same material (61~meV)~\cite{xu2025origin}. For $T_{\text{DW}}=140$~K, this yields $2\Delta/k_B T_{\text{DW}}\approx 9.3$, far above the weak-coupling value of 3.52, highlighting the unconventional nature of the DW transition in La$_4$Ni$_3$O$_{10}$.

The temperature dependence of energy gap $\Delta$ [Fig.~\ref{fig:S4_energy_gap_MF}(c)] follows a mean-field behavior. We obtain the mean-field curve by solving the self-consistent BCS gap equation and normalizing by the maximum gap and the DW transition temperature. To estimate uncertainties, we conservatively assume a relative reflectivity error of $\pm0.5\%$ between temperatures, rescale $R(\omega)$ at all temperatures by factors of 1.005 and 0.995, recompute the corresponding conductivities, and re-extract $\Delta$; we take the difference between the gap values from the scaled reflectivities as the error bar, providing a conservative estimate of uncertainty from reflectivity errors. While the number of data points near $T_{\text{DW}}$ is limited, the observed $\Delta(T)$ suggests that the DW transition is close to continuous and likely driven by an SDW instability, consistent with the Landau analysis of Ref.~\cite{norman2025landau}.

\section{Electronic correlation calculation} \label{sec:correlation}

We quantify electronic correlations in La$_4$Ni$_3$O$_{10}$ using the ratio of experimental to band-theory kinetic energy, $K_{\mathrm{exp}}/K_{\mathrm{DFT}}$. To avoid contributions from interband transitions and phonons, we relate the kinetic energy to the Drude plasma frequency and compute $K_{\text{exp}}/K_{\text{DFT}}=\omega_{p,exp}^2/\omega_{p,cal}^2$. Because standard DFT does not capture DW effects, we use the experimental optical conductivity at 160~K (just above the DW transition) as the reference experimental spectrum. We perform the DFT calculation without including a Hubbard $U$. For the in-plane direction, a Drude$+\varepsilon_\infty$ fit to the measured $\sigma_{1,ab}(\omega)$ yields $\omega_{p,\mathrm{exp}}^{ab}=9855$~cm$^{-1}$ (1.222~eV), while DFT gives $\omega_{p,\mathrm{cal}}^{ab}=2.511$~eV, resulting in $K^{ab}_{\mathrm{exp}}/K^{ab}_{\mathrm{DFT}}=0.237$, consistent with previous reports~\cite{xu2025origin, liu2025evolution}. For the out-of-plane direction, we obtain $\omega_{p,\mathrm{exp}}^{c^*}=402$~cm$^{-1}$ (0.050~eV) and $\omega_{p,\mathrm{cal}}^{c^*}=0.278$~eV, giving $K^{c^*}_{\mathrm{exp}}/K^{c^*}_{\mathrm{DFT}}=0.032$.

\begin{figure}
    \centering
    \includegraphics[width=0.6\textwidth]{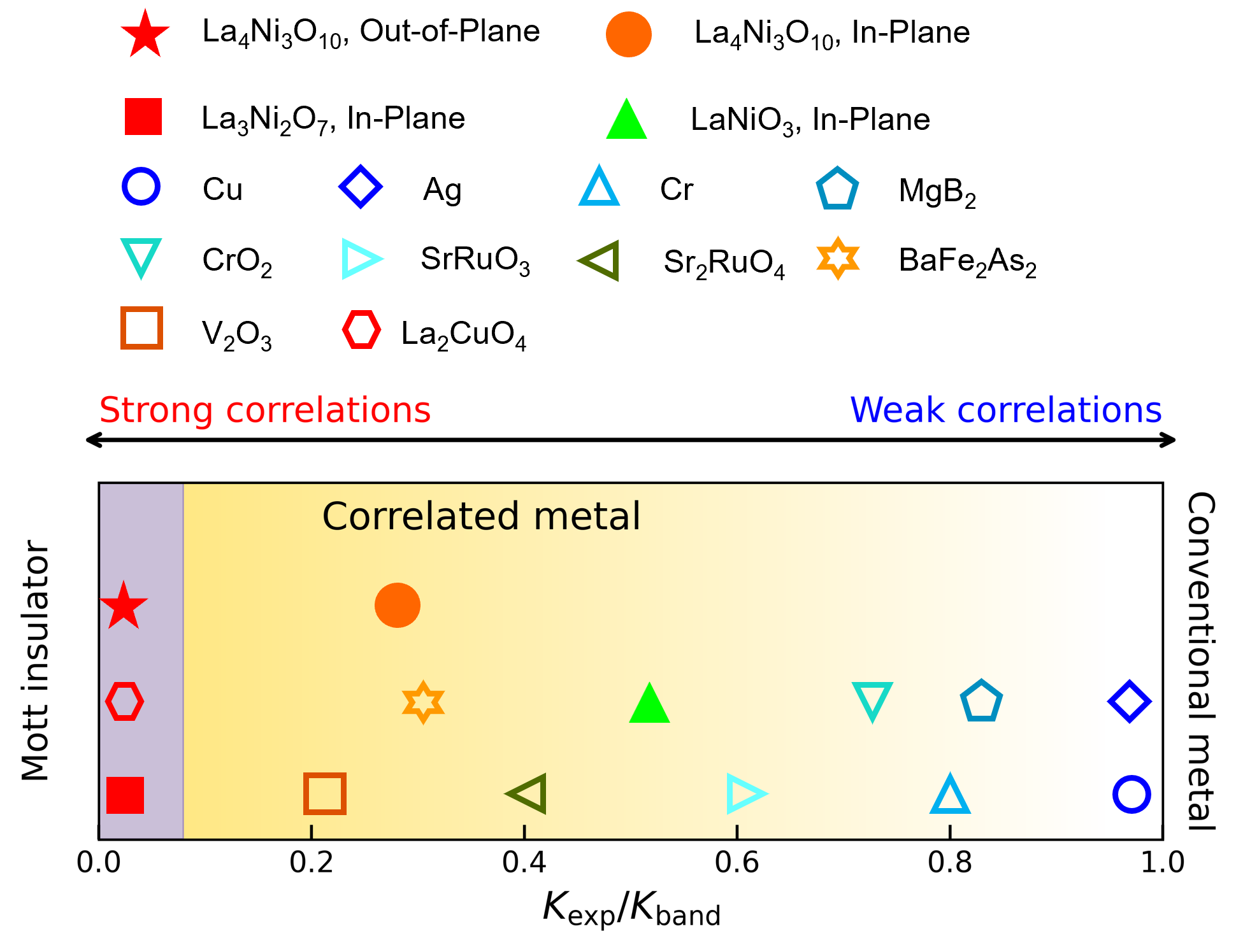}
    \caption{Kinetic-energy ratio $K_{\mathrm{exp}}/K_{\mathrm{band}}$ for La$_4$Ni$_3$O$_{10}$ along the in-plane and out-of-plane directions, compared with the related nickelates La$_3$Ni$_2$O$_7$ and LaNiO$_3$, and with other representative materials. For clarity, nickelates are shown as solid symbols and other materials as open symbols. La$_4$Ni$_3$O$_{10}$ is highlighted by a solid red star (out of plane) and a solid orange circle (in plane). Values of $K_{\mathrm{exp}}/K_{\mathrm{band}}$ are taken from Refs.~\cite{qazilbash2009electronic, liu2025evolution, liu2024electronic} and references therein.}
    \label{fig:S5_correlation_compare}
\end{figure}

Based on kinetic-energy ratios extracted from the optical plasma frequencies, La$_4$Ni$_3$O$_{10}$ is a moderately correlated metal in the in-plane direction, with $K^{ab}_{\mathrm{exp}}/K^{ab}_{\mathrm{DFT}}=0.237$, while the out-of-plane response is strongly suppressed, yielding $K^{c^*}_{\mathrm{exp}}/K^{c^*}_{\mathrm{DFT}}=0.032$. As summarized in Fig.~\ref{fig:S5_correlation_compare}, these values place La$_4$Ni$_3$O$_{10}$ in an intermediate regime relative to other layered nickelates and correlated oxides. The in-plane response is comparable to BaFe$_2$As$_2$, where multiorbital physics is important, whereas the out-of-plane response resembles La$_3$Ni$_2$O$_7$ and La$_2$CuO$_4$, which lie close to the Mott-insulating limit. This pronounced anisotropy reflects the layered crystal structure and the strongly reduced interlayer charge dynamics.

\section{Drude-Lorentz Fitting}\label{sec:DLfit}
We use the multi-component Drude-Lorentz model to quantitatively analyze the temperature dependence of free carrier density, scattering rate, and phonon modes. The optical response of the material is modeled using a standard Drude–Lorentz formalism for the dimensionless complex dielectric function~\cite{dressel2002electrodynamics,tanner2015use},

\begin{equation}\label{eq:DLmodel}
\varepsilon(\omega) = \varepsilon_\infty
- \sum_{j} \frac{\omega_{p,j}^2}{\omega^2 + i\gamma_j \omega}
+ \sum_{k} \frac{\omega_{p,k}^2}{\omega_{0,k}^2 - \omega^2 - i\gamma_k \omega},
\end{equation}
where the first summation describes the free-carrier response using one or more Drude terms, characterized by Drude plasma frequencies $\omega_{p,j}$ and Drude scattering rates $\gamma_j$. The second summation represents bound excitations modeled as Lorentz oscillators with resonance frequencies $\omega_{0, k}$, oscillator strengths $\omega_{p,k}$, and damping constants $\gamma_k$, which captures infrared-active phonons, interband transitions, and other finite energy excitations. The term $\epsilon_\infty$ is the high-frequency dielectric constant accounting for polarization processes that occur at energies well above the spectral window explicitly modeled by the Drude and Lorentz terms. Physically, $\epsilon_\infty$ incorporates mostly the cumulative contribution of core-electron and high-energy valence-electron screening that respond essentially instantaneously on the timescale of infrared excitations. As such, $\epsilon_\infty$ provides a constant background that ensures the correct high-frequency limit of the dielectric function and should not be interpreted as the static dielectric constant.

The complex optical conductivity $\sigma(\omega)$ is directly related to the dielectric function via
\begin{equation}
\sigma(\omega) = \frac{\omega \left[ \varepsilon(\omega) - 1 \right]}{59.958\, i},
\end{equation}
where the frequency $\omega$ is expressed in units of cm$^{-1}$ and the resulting conductivity is obtained in units of $\Omega^{-1}$cm$^{-1}$. The numerical factor $59.958\approx60$ arises from unit conversion and corresponds to $Z_0/2\pi$, where $Z_0$ is the vacuum impedance~\cite{dressel2002electrodynamics,tanner2015use}. We use this formula to simultaneously fit the real and imaginary parts of the complex optical conductivity. 

\begin{figure}
    \centering
    \includegraphics[width=0.98\textwidth]{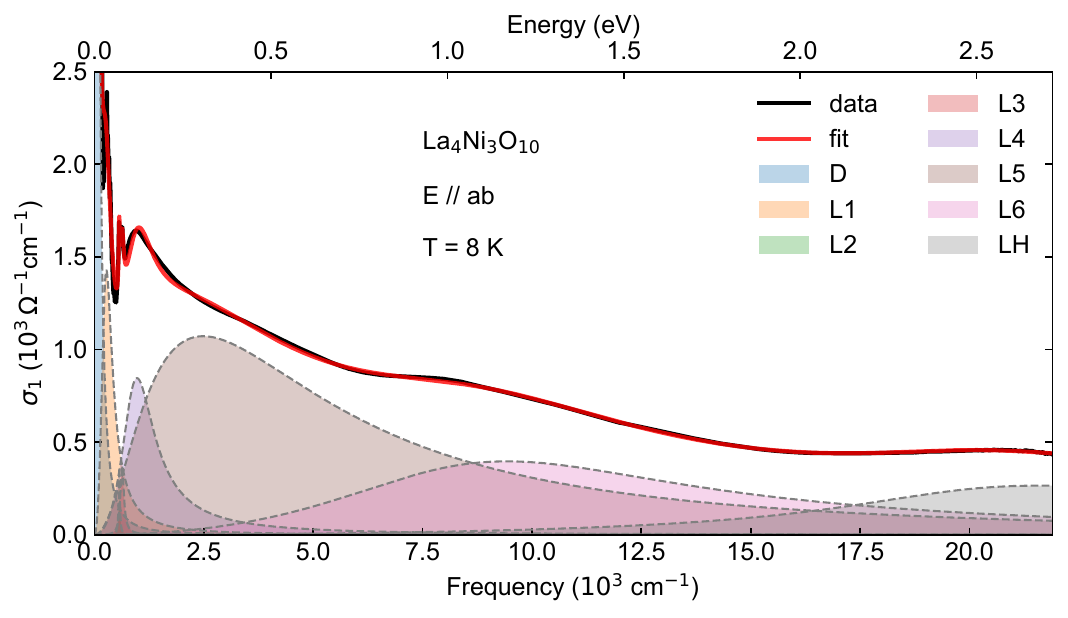}
    \caption{The in-plane experimental $\sigma_1(\omega)$ (black curve) at 8\,K and the Drude-Lorentz fitting result (red curve). Individual Drude and Lorentz components are shown as shaded areas.}
    \label{fig:S7_abfit_full}
\end{figure}

Fig.~\ref{fig:S7_abfit_full} shows the Drude-Lorentz fit to the in-plane optical conductivity at 8~K. We model multiple interband transitions with a set of Lorentz oscillators and include a phonon-like feature near 600~cm$^{-1}$. The fitted interband structure broadly follows DFT, although the oscillator strengths differ quantitatively, and fewer distinct modes are resolved experimentally, consistent with substantial spectral broadening.

In RP nickelates, the low-frequency response is sometimes parameterized with multiple Drude components to reflect multiband character and possible orbital-dependent scattering~\cite{liu2024electronic,xu2025origin}. When these intraband contributions are not spectroscopically resolved, however, such decompositions are not unique and increase correlations among parameters. Here, the low-frequency conductivity is a single, smooth Drude-like component, without clear signatures that uniquely require multiple Drude terms. Hence, we adopt a minimal intraband description to capture the essential charge dynamics while limiting model dependence.

\begin{figure}
    \centering
    \includegraphics[width=0.98\textwidth]{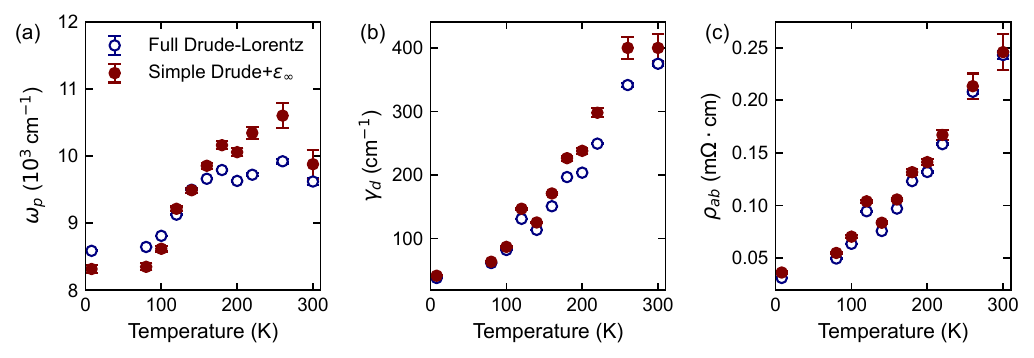}
    \caption{Comparison of (a) in-plane Drude plasma frequency $\omega_p$, (b) scattering rate $\gamma_d$, and (c) DC resistivity $\rho_{ab}$  obtained from full-range Drude-Lorentz fitting (blue open circles) and low-frequency Drude+$\varepsilon_\infty$ analysis (red filled circles) and as function of temperature.}
    \label{fig:S8_full_simple_fit_compare}
\end{figure}

We also compare two fitting strategies for the in-plane response in Fig.~\ref{fig:S8_full_simple_fit_compare}: (i) a full-range Drude-Lorentz fit over the entire measured window and (ii) a simple single Drude+$\varepsilon_\infty$ fit. The extracted plasma frequency, scattering rate, and DC resistivity from the two approaches agree within uncertainties, supporting the Drude$+\varepsilon_\infty$ analysis used in the main text as a reliable description of the in-plane free-carrier response.

For the out-of-plane response, the Drude spectral weight is much smaller, so full-range Drude-Lorentz fits become dominated by higher-frequency contributions. The resulting low-energy Drude parameters depend sensitively on the high-frequency modeling and are therefore not robust. We thus avoid full-range Drude-Lorentz fits for the $c$-axis data and instead rely on a low-frequency Drude$+\varepsilon_\infty$ analysis.

We observe that the in-plane response the squared Drude plasma frequency $\omega_p^2$, which is proportional to the effective free-carrier density, is reduced by approximately 25\% upon cooling from high temperature to low temperature. This suppression indicates a partial depletion of the in-plane Fermi surface, consistent with the opening of a density-wave–induced gap that removes a fraction of the itinerant carriers~\cite{xu2025origin}.

\begin{figure}
    \centering
    \includegraphics[width=0.98\textwidth]{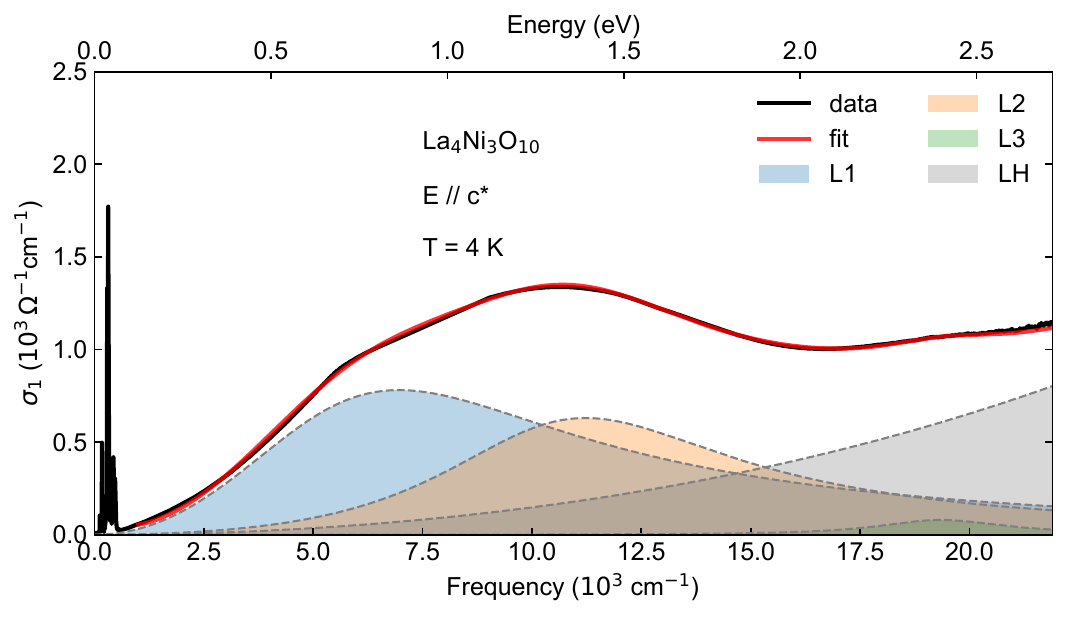}
    \caption{Out-of-plane experimental $\sigma_1(\omega)$ (black curve) at 4\,K and the Drude-Lorentz fit (red curve). Individual Lorentz components are shown as shaded areas.}
    \label{fig:S9_cfit_high}
\end{figure}

Fig.~\ref{fig:S9_cfit_high} shows the Drude-Lorentz fit to the out-of-plane optical conductivity at 4~K. We model high-energy interband transitions ($>1000$~cm$^{-1}$) with a set of Lorentz oscillators, while we defer the detailed Drude-Lorentz analysis of phonons below 600~cm$^{-1}$ to Sec.~\ref{sec:PhononFit}. As for the in-plane response, the fitted interband structure broadly agrees with DFT, although the oscillator strengths differ quantitatively and fewer modes are resolved experimentally, consistent with substantial spectral broadening.

\section{Transport Resistivity}\label{sec:Transport}
To further validate the temperature-dependent anisotropy extracted from the optical conductivity, we compare the DC resistivity inferred from optics with transport measurements. Transport resistivity carries inherent systematic uncertainties, most notably geometric factors used to convert resistance to resistivity, and overall uncertainties can approach an order of magnitude. A major advantage of optical spectroscopy is to provide low-frequency limit resistivity free from the geometric uncertainties. For this reason, in comparing to transport we apply an overall rescaling factor to the transport data to account for its absolute uncertainty.

\begin{figure}
    \centering
    \includegraphics[width=0.6\textwidth]{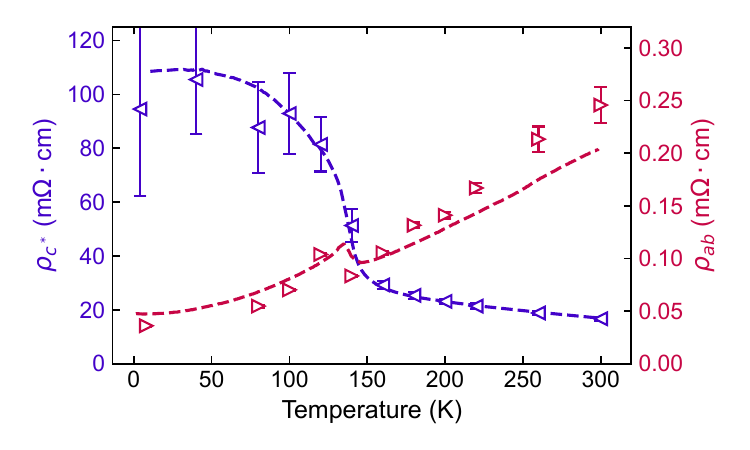}
    \caption{Comparison between DC resistivity extracted from optical (triangles with error bars) and transport measurements (broken lines) on La$_4$Ni$_3$O$_{10}$ along both in-plane and out-of-plane directions. The in-plane transport resistivity $\rho_{ab}$ data are adapted and rescaled from Ref. \cite{zhang2020high}.}
    \label{fig:S6_transport_compare}
\end{figure}

For the in-plane resistivity, we use published transport data from Ref.~\cite{zhang2020high}, shown as the cardinal red dashed line in Fig.~\ref{fig:S6_transport_compare} after applying a uniform rescaling factor of 0.15. Since no out-of-plane resistivity has been reported for La$_4$Ni$_3$O$_{10}$, we measured $\rho_{c^*}$ using a standard four-probe technique on a polished $ac^*$-plane sample, varying temperature from 10~K to 300~K in a He$^4$ gas-flow cryostat. The measured out-of-plane $\rho_{c^*}(T)$ is also rescaled by a constant factor for plotting in Fig.~\ref{fig:S6_transport_compare}, to align with the optically extrapolated DC resistivity. We note that in highly anisotropic systems such as La$_4$Ni$_3$O$_{10}$ it is difficult to eliminate mixing of in- and out-of-plane channels in a conventional four-probe geometry, and some sample-to-sample dispersion is also expected~\cite{zhang2020high, rout2020structural, kumar2020physical}.
With these caveats in mind, the temperature dependence of our measured $\rho_{c^*}(T)$ closely resembles the reported out-of-plane resistivity trend of Pr$_4$Ni$_3$O$_{10}$ in Ref.~\cite{huangfu2020anisotropic}. Overall, the in-plane and out-of-plane resistivities inferred from the optical Drude analysis are in reasonable agreement with the transport data at the level of temperature trends.
We emphasize that, due to experimental schematics and sample-to-sample dispersion, the absolute value of $\rho_{c^*}/\rho_{ab}$ should be taken as an estimate, while the temperature trend is the more robust result.

\section{Phonon Modes: Fitting}\label{sec:PhononFit}
\begin{figure}
    \centering
    \includegraphics[width=0.9\textwidth]{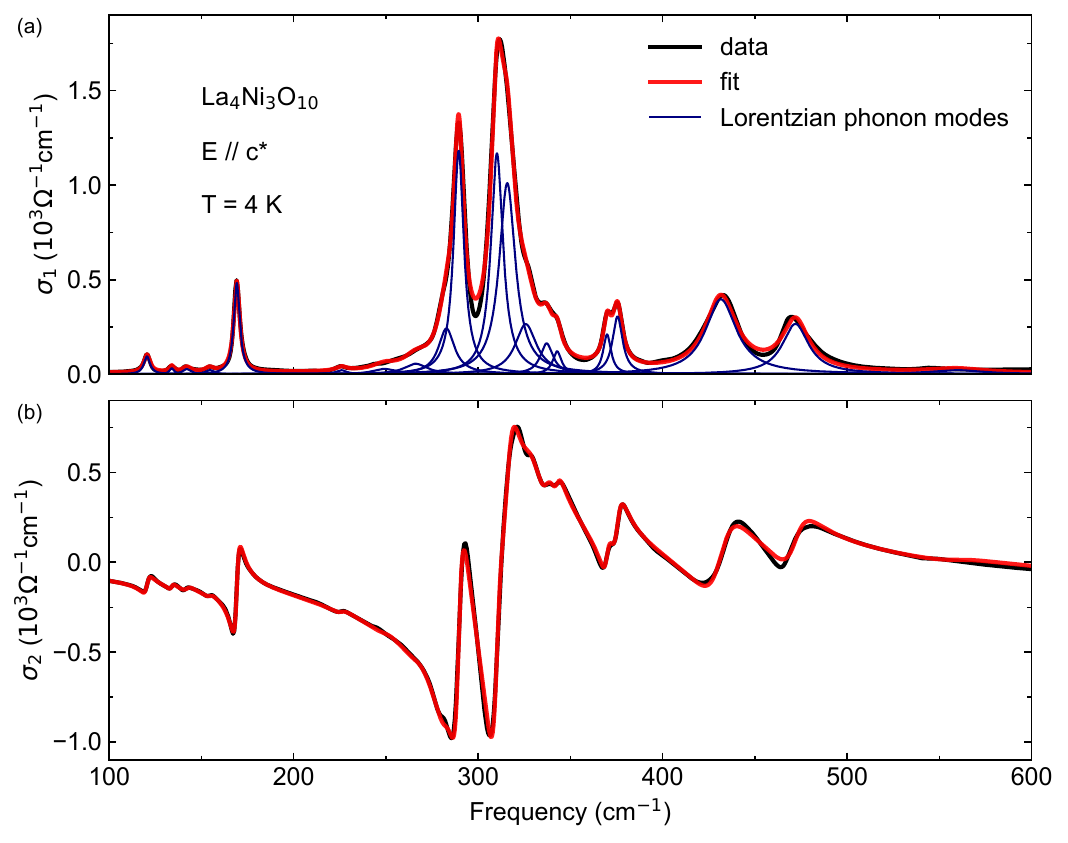}
    \caption{Real and imaginary optical conductivity $\sigma_1(\omega)$ and $\sigma_2(\omega)$ at 4\,K in the range of 100-600~cm$^{-1}$. The solid red line is the Drude-Lorentz fit of the experimental data (black solid line). Thin navy solid lines correspond to individual Lorentzian phonon modes. We identify a total of 20 modes at 4\,K.}
    \label{fig:S10_DLphonon}
\end{figure}

\begin{figure}
    \centering
    \includegraphics[width=0.9\textwidth]{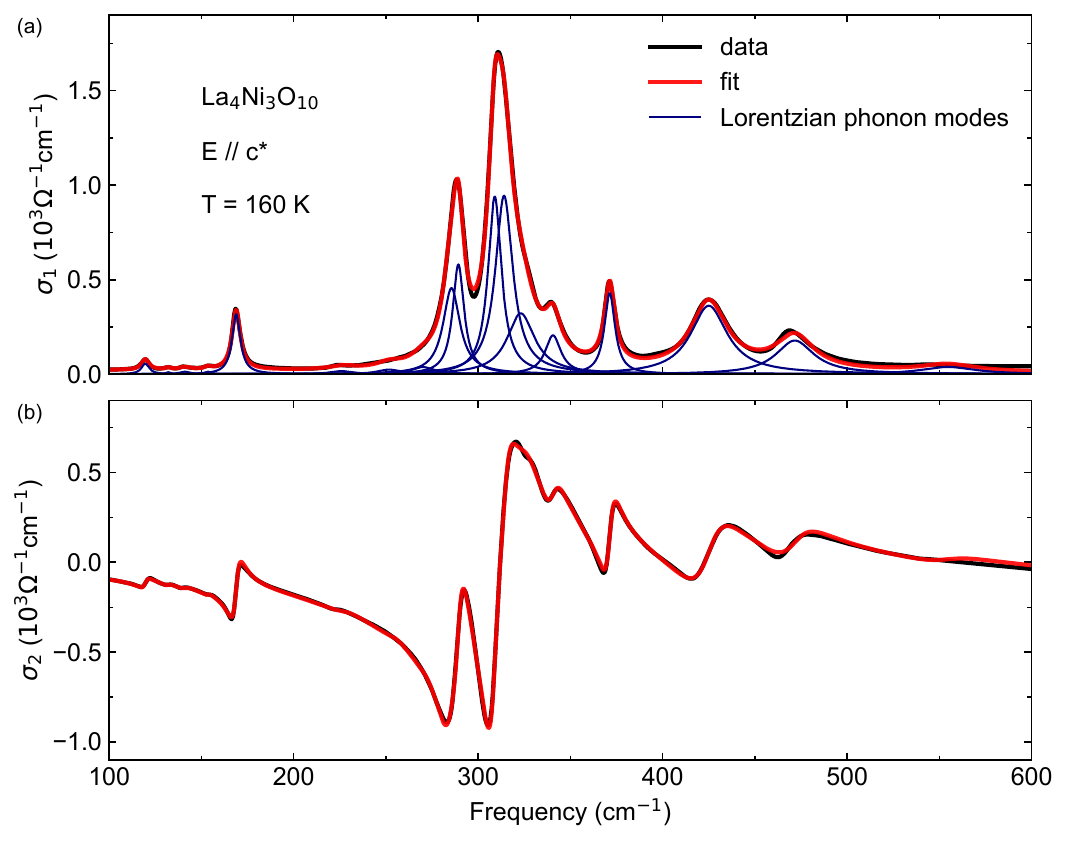}
    \caption{Real and imaginary optical conductivity $\sigma_1(\omega)$ and $\sigma_2(\omega)$ at 160\,K in the range of 100-600~cm$^{-1}$. The solid red line is the Drude-Lorentz fit of the experimental data (black solid line). Thin navy solid lines correspond to individual Lorentzian phonon modes. We identify total of 18 modes at 160\,K.}
    \label{fig:S11_DLphonon160}
\end{figure}

We fit the $c^*$-axis optical conductivity with a Drude-Lorentz model over the full temperature range, restricting the window to $\omega<600$~cm$^{-1}$ to include all observed infrared-active phonons while excluding higher-energy interband transitions. Fits at 4~K (Fig.~\ref{fig:S10_DLphonon}) and 160~K (Fig.~\ref{fig:S11_DLphonon160}) are shown as representative examples well below and above the DW transition, respectively. In both cases, simultaneous fits to the real and imaginary parts of the complex conductivity capture the main features, enabling a systematic analysis of the phonon evolution with temperature.

\section{Phonon Modes: Comparison with DFT}\label{sec:PhononDFT}
At ambient-pressure, La$_4$Ni$_3$O$_{10}$ crystallizes in the monoclinic $P2_1/a$ structure and the structural symmetry is preserved across the DW transition \cite{kumar2020physical, rout2020structural, li2025crystal}. This structure with two formula units per primitive unit cell ($Z=2$) supports a total of $102$ normal modes at the Brillouin zone center: 3 acoustic and 99 optical modes. These modes decompose into $24A_g+26A_u+24B_g+25B_u$ representations, where the odd-parity $A_u$ and $B_u$ modes are infrared-active and the even-parity $A_g$ and $B_g$ modes are Raman-active. 
In the conventional monoclinic $P2_1/a$ setting (unique $b$ axis), the infrared-active modes decompose into $A_u$ modes with dipole moments primarily along $b$, and $B_u$ modes with dipole in the $ac$ plane. Thus, an out-of-plane polarized experiment is expected to predominantly probe 
$B_u$ modes. The monoclinic tilt angle ($\beta\approx100.8^\circ$, i.e. $\sim11^\circ$ between the crystallographic $c$ direction and the experimental out-of-plane $c^*$ direction~\cite{zhang2020high})  together with crystal twinning effects and a small (possible) miscut can mix components, so both $A_u$ and $B_u$ modes may acquire finite intensity for the out-of-plane measurement.

Given the large number of symmetry-allowed optical modes and the monoclinic symmetry, we use first-principles calculations to guide the identification of corresponding modes. We perform phonon calculations within density functional theory using the finite-displacement method. We carry out the electronic-structure calculations with the Vienna \textit{ab initio} Simulation Package (VASP) using the GGA-PBE exchange-correlation functional~\cite{gga_pbe}.
We base the calculations on the experimentally determined monoclinic $P2_1/a$ structure of La$_4$Ni$_3$O$_{10}$, with structural data taken from Ref. ~\cite{zhang2020high}. We obtain interatomic force constants with \textsc{phonopy}\cite{phonopy} by finite displacements in a $2\times2\times1$ supercell, corresponding to a doubling of the in-plane unit cell. We then use these force constants to compute zone-center phonon frequencies and eigenvectors.

Table~\ref{tab:Au_phonon_160K_4K} compares the DFT zone-center frequencies of infrared-active phonons with the mode frequencies extracted from Drude-Lorentz fits at 160~K and 4~K. At 160~K, above the density-wave transition, we associate the fitted out-of-plane modes with nearby calculated infrared-active modes based on frequency proximity and the calculated oscillator strengths of the out-of-plane dipole component. By contrast, at 4~K (below $T_\text{DW}$) we resolve additional phonon modes that have no direct counterparts among the calculated modes.

\begin{table}[h]
\centering
\caption{Comparison of extracted phonon modes from Drude-Lorentz fits at 160~K and 4~K to calculated infrared-active phonon modes. The DFT mode labels ($P\#$) follow the ranking of the phonon frequencies among all 102 calculated modes. }
\label{tab:Au_phonon_160K_4K}
\begin{tabular}{
c@{\hspace{12pt}}
c@{\hspace{14pt}}
c@{\hspace{14pt}}
c
}
\hline \hline
\textbf{DFT modes} & \textbf{DFT (cm$^{-1}$)} & \textbf{160~K fit (cm$^{-1}$)} & \textbf{4~K fit (cm$^{-1}$)} \\
\hline \hline
\noalign{\global\setlength{\arrayrulewidth}{0.3pt}}

$P18$  & 120.6 & $119.6 \pm 0.1$ & $120.5 \pm 0.1$ \\ \hline
$P20$  & 125.5 & $132.0 \pm 0.3$ & $133.9 \pm 0.2$ \\ \hline
$P23$  & 137.3 & $140.9 \pm 0.4$ & $142.0 \pm 0.3$ \\ \hline
$P26$  & 144.2 & $153.5 \pm 0.3$ & $154.4 \pm 0.3$ \\ \hline
$P28$  & 152.1 & $168.8 \pm <0.1$ & $169.2 \pm <0.1$ \\ \hline

$P46$  & 227.5 & $225.9 \pm 0.8$ & $225.9 \pm 0.6$ \\ \hline
$P48$  & 234.0 & $251.6 \pm 0.7$ & $249.2 \pm 1.1$ \\ \hline
$P54$  & 268.3 & $270.0 \pm 0.7$ & $266.1 \pm 0.6$ \\ \hline
$P57$  & 286.2 & $285.6 \pm 0.4$ & $282.8 \pm 0.2$ \\ \hline
$P60$  & 296.0 & $289.3 \pm 0.1$ & $289.5 \pm <0.1$ \\ \hline
$P62$  & 303.8 & $309.0 \pm 0.1$ & $310.2 \pm 0.1$ \\ \hline
$P64$  & 307.6 & $314.0 \pm 0.2$ & $315.8 \pm 0.1$ \\ \hline
$P68$  & 316.4 & $323.1 \pm 0.6$ & $325.8 \pm 0.4$ \\ \hline
\multirow{2}{*}{$P76$} & \multirow{2}{*}{353.3} & \multirow{2}{*}{$340.6 \pm 0.1$} & $337.1 \pm 0.3$ \\
\thincline{4-4}
                       &                         &                               & $342.8 \pm 0.3$ \\ \hline

\multirow{2}{*}{$P81$} & \multirow{2}{*}{390.9} & \multirow{2}{*}{$371.3 \pm <0.1$} & $369.9 \pm 0.1$ \\
\thincline{4-4}
                       &                         &                                  & $375.6 \pm 0.1$ \\ \hline

$P86$  & 434.4 & $425.0 \pm 0.1$ & $431.6 \pm 0.1$ \\ \hline
$P93$  & 501.1 & $471.6 \pm 0.2$ & $472.0 \pm 0.2$ \\ \hline
$P100$  & 573.9 & $554.5 \pm 1.2$ & $558.9 \pm 3.6$ \\ \hline
\end{tabular}
\end{table}

\section{Phonon Modes: Anomaly and atomic displacement}\label{sec:phononAnomalyDisplace}
Across the DW transition, reported lattice-parameter anomalies are well below the level of 0.1\% in relative magnitude~\cite{zhang2020high, kumar2020physical, rout2020structural, li2025crystal}. A simple way to estimate the size of phonon shifts expected from such small structural changes is via a mode Gr\"uneisen parameter \(\gamma_i\), defined by
\begin{equation}
\frac{\Delta \omega_i}{\omega_i}\approx -\gamma_i\,\frac{\Delta V}{V},
\end{equation}
(or an anisotropic analog using linear strains). The Gr\"uneisen parameter, \(\gamma_i\), is usually of order unity for common solid-state materials, even taking a conservative upper bound \(\gamma_i\sim 2\text{-}3\)~\cite{boehler1980experimental, bruls2001temperature}, lattice/volume changes below 0.1\% level would imply \(\Delta\omega/\omega \lesssim 0.1\%\text{-}0.3\%\). This estimate is well below the \(\sim 1\%\) frequency anomalies and splittings we observe, and, importantly, such anomalies are highly mode-selective (confined to the 300-500~cm\(^{-1}\) window) rather than a broadly distributed hardening/softening expected from uniform strain. This comparison supports that the dominant phonon renormalization is tied to the DW electron-phonon/magnetoelastic coupling rather than a simply structural anomaly.

To quantify the temperature dependence of the phonon anomalies, we analyze the evolution of phonon frequency $\omega_0(T)$ and scattering rate (linewidth) $\gamma(T)$ extracted from the Drude-Lorentz fits. The high-temperature behavior is described by the Klemens-Hart-Aggarwal-Lax model \cite{balkanski1983anharmonic}. In this model, cubic anharmonic decay of phonons gives the temperature dependence of the phonon frequency and the scattering rate as
\begin{equation} \label{eq:anharmonicphonon}
\begin{aligned}
\omega_0(T) &= \omega_{0, T=0} + A \cdot \left(1 + \frac{2}{e^{x} - 1} \right), \\
\gamma(T) &= \gamma_{\text{res}} + B\cdot \left(1 + \frac{2}{e^{x} - 1} \right),
\end{aligned}
\end{equation}

with $x = \frac{\hbar \omega_{0,T=0}}{2 k_B T}$. 

Here, $\omega_{0,T=0}$ denotes the bare phonon frequency at zero temperature, $\gamma_{\mathrm{res}}$ represents a temperature-independent residual scattering rate that captures additional broadening mechanisms not accounted for by anharmonic decay, and $A, B$ are constants. Fits using these expressions are restricted to the high-temperature regime, and deviations from the extrapolated behavior at low temperatures reflect strong phonon renormalization induced by the DW transition.

\begin{figure}
    \centering
    \includegraphics[width=0.98\textwidth]{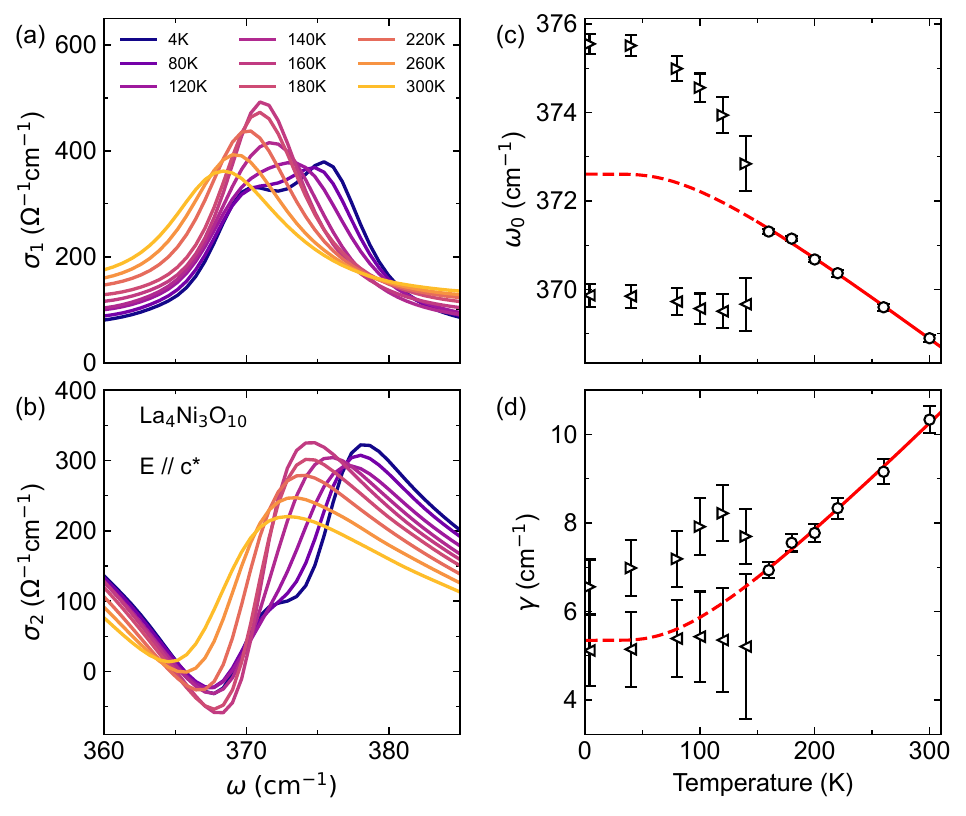}
    \caption{(a), (b) Real and imaginary out-of-plane optical conductivity of La$_4$Ni$_3$O$_{10}$ in a narrow frequency window around the 371~cm$^{-1}$ mode. (c), (d) Temperature dependence of the fitted phonon frequency $\omega_0$ and scattering rate $\gamma$ for the modes near near 371~cm$^{-1}$. On cooling below the density-wave (DW) transition, this phonon splits into two distinct low-temperature branches that evolve from a single high-temperature mode. Solid red curves show anharmonic phonon-decay fits to the high-temperature data, and dashed curves extrapolate the fits to low temperature. The strong deviation of the measured low-temperature phonon parameters from the extrapolated anharmonic trend indicates substantial DW-associated phonon renormalization.}
    \label{fig:S12_P375_split}
\end{figure}

\begin{figure}
    \centering
    \includegraphics[width=0.98\textwidth]{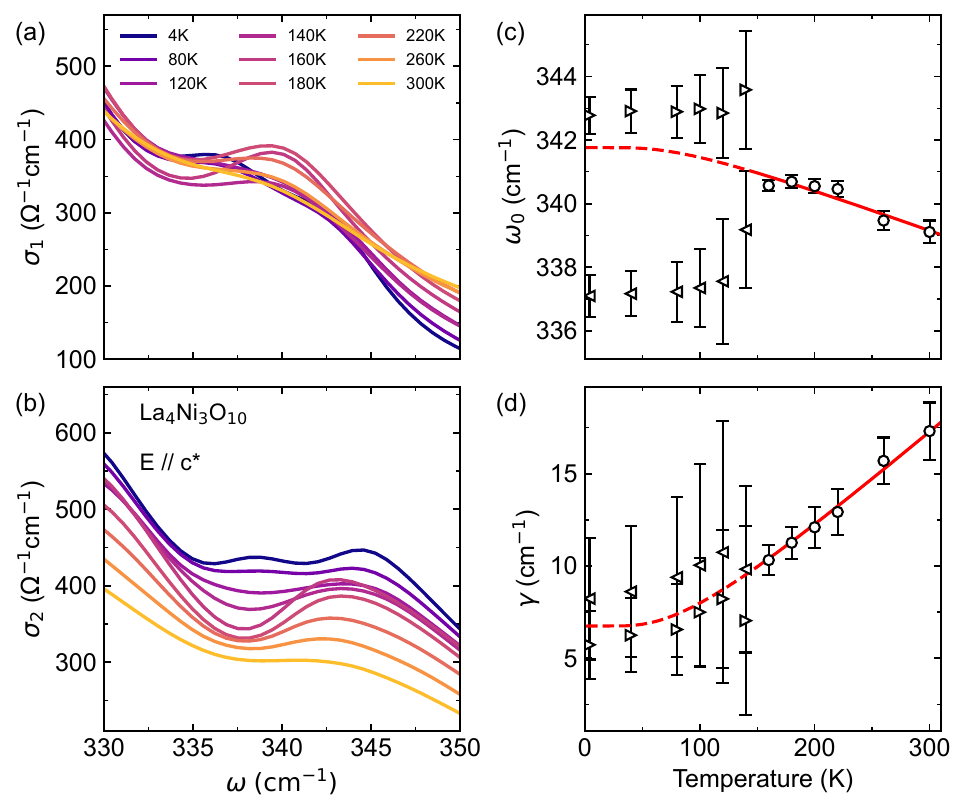}
    \caption{(a), (b) Real and imaginary out-of-plane optical conductivity of La$_4$Ni$_3$O$_{10}$ in a narrow frequency window around the 340~cm$^{-1}$ mode. (c), (d) Temperature dependence of the fitted phonon frequency $\omega_0$ and scattering rate $\gamma$ for the modes near near 340~cm$^{-1}$. The phonon mode exhibits a subtle splitting below the DW transition, most clearly resolved in the imaginary part of the optical conductivity $\sigma_2(\omega)$. Solid red curves represent fits to the high-temperature data using an anharmonic phonon decay model, whereas dashed lines indicate extrapolations into the low-temperature regime. The deviation between the low-temperature phonon parameters and the extrapolated anharmonic behavior reflects a DW-induced phonon renormalization.
}
    \label{fig:S13_P340_split}
\end{figure}

\begin{figure}
    \centering
    \includegraphics[width=0.98\textwidth]{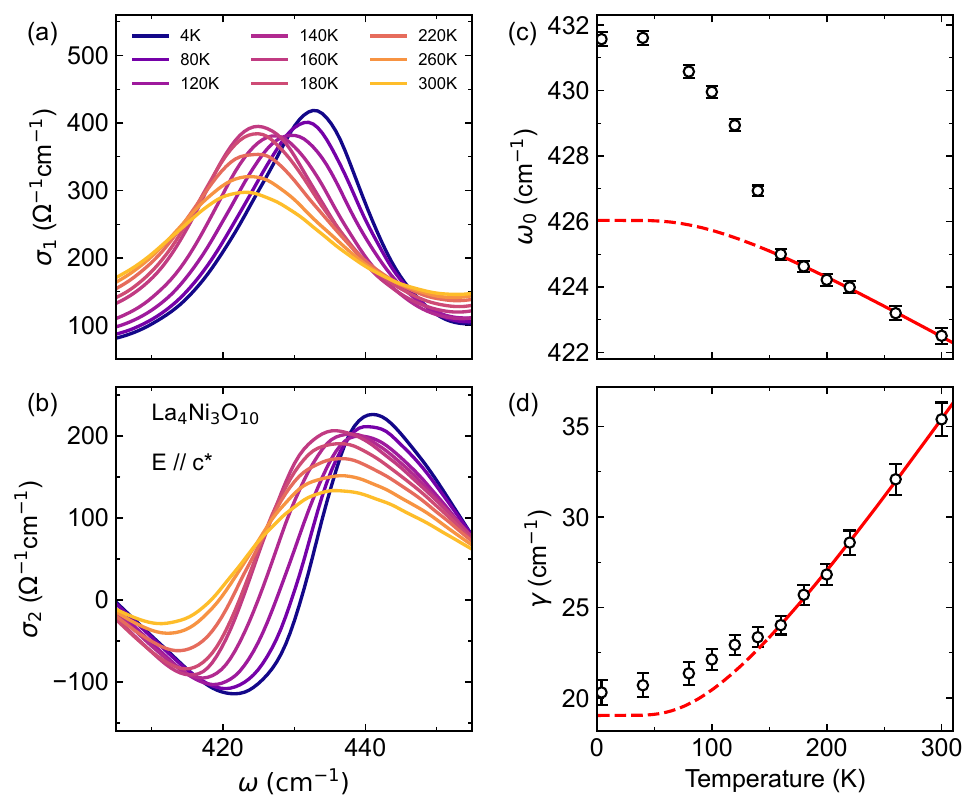}
    \caption{(a), (b) Real and imaginary out-of-plane optical conductivity of La$_4$Ni$_3$O$_{10}$ in a narrow frequency window around the 430~cm$^{-1}$ mode. (c), (d) Temperature dependence of the fitted phonon frequency $\omega_0$ and scattering rate $\gamma$ for the modes near near 430~cm$^{-1}$.Unlike the lower-energy modes, this phonon shows no resolvable splitting, but its frequency and linewidth renormalize strongly on cooling below the DW transition. Solid red curves show anharmonic phonon-decay fits to the high-temperature data, and dashed curves extrapolate the fits to low temperature. The marked deviation of the low-temperature phonon parameters from the extrapolated anharmonic trend signals strong DW-induced renormalization.}
    \label{fig:S14_P430_renorm}
\end{figure}

\begin{figure}
    \centering
    \includegraphics[width=0.98\textwidth]{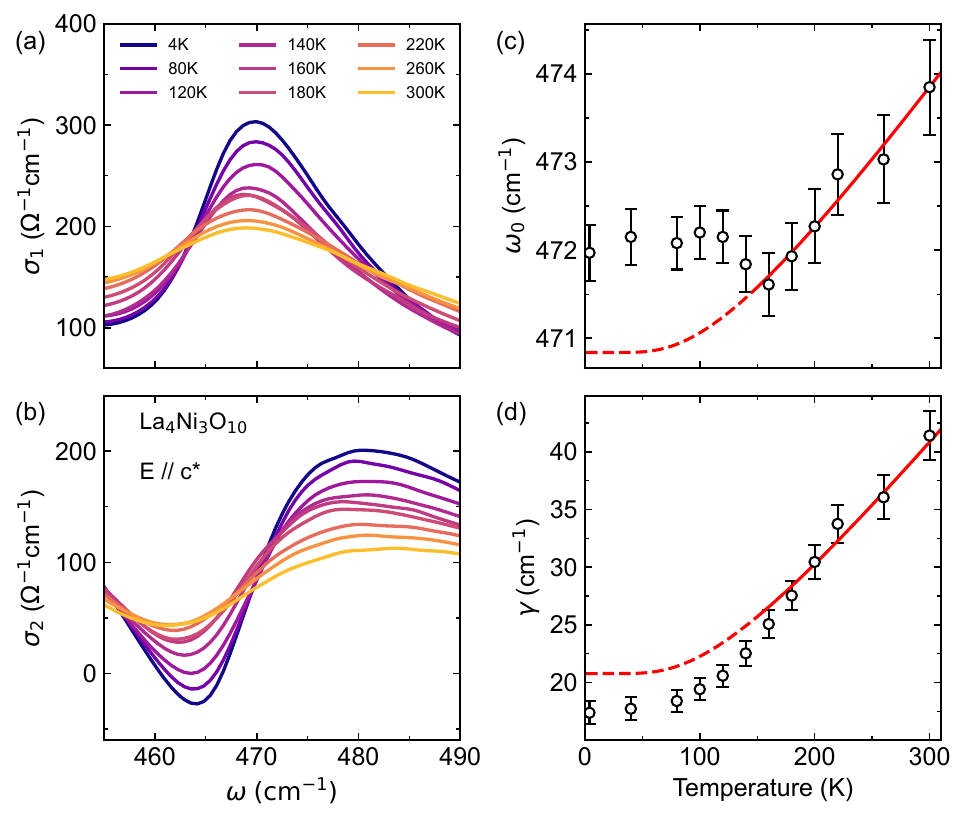}
    \caption{(a), (b) Real and imaginary out-of-plane optical conductivity of La$_4$Ni$_3$O$_{10}$ in a narrow frequency window around the 475~cm$^{-1}$ mode. (c), (d) Temperature dependence of the fitted phonon frequency $\omega_0$ and scattering rate $\gamma$ for the modes near near 475~cm$^{-1}$. As for the mode near 430~cm$^{-1}$, this phonon shows no clear splitting but undergoes a strong renormalization of its frequency and linewidth below the DW transition. Solid red curves show anharmonic phonon-decay fits to the high-temperature data, and dashed curves extrapolate the fits to low temperature. The deviation of the measured low-temperature parameters from the extrapolated anharmonic trend further highlights the strong impact of the DW transition on the lattice dynamics.}
    \label{fig:S15_P475_renorm}
\end{figure}

\begin{figure}
    \centering
    \includegraphics[width=0.98\textwidth]{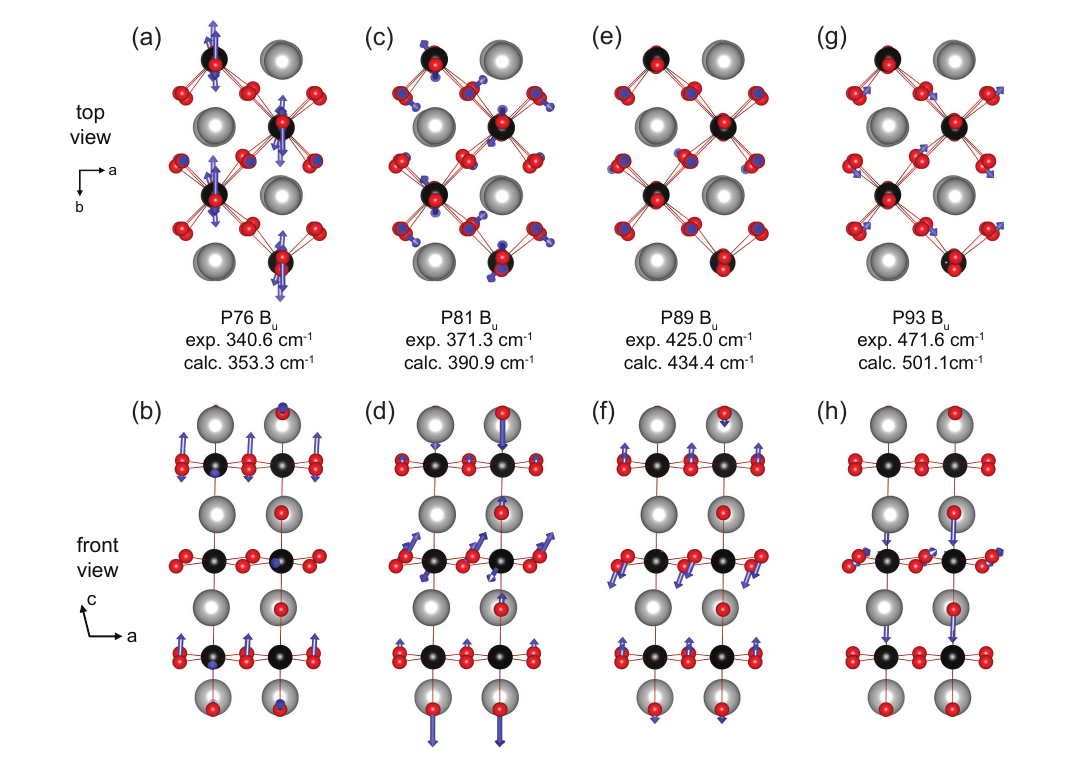}
    \caption{Calculated atomic displacement patterns for four representative infrared-active phonon modes that exhibit pronounced anomalies across the density-wave transition. Panels (a,b), (c,d), (e,f), and (g,h) show the top and front views of the phonon eigenvectors for modes near 340~cm$^{-1}$ (P76), 371~cm$^{-1}$ (P81), 430~cm$^{-1}$ (P86), and 475~cm$^{-1}$ (P93), respectively. The corresponding experimental (at 160~K) and calculated phonon frequencies are indicated for each mode. The mode labels (P\#) follow the ranking of the phonon frequencies among all 102 calculated zone-center phonon modes, consistent with the notation used in Table \ref{tab:Au_phonon_160K_4K}. For clarity, only displacements with large amplitude are shown for each mode. }
    \label{fig:S16_more_diplacement}
\end{figure}

\begin{figure}
    \centering
    \includegraphics[width=0.98\textwidth]{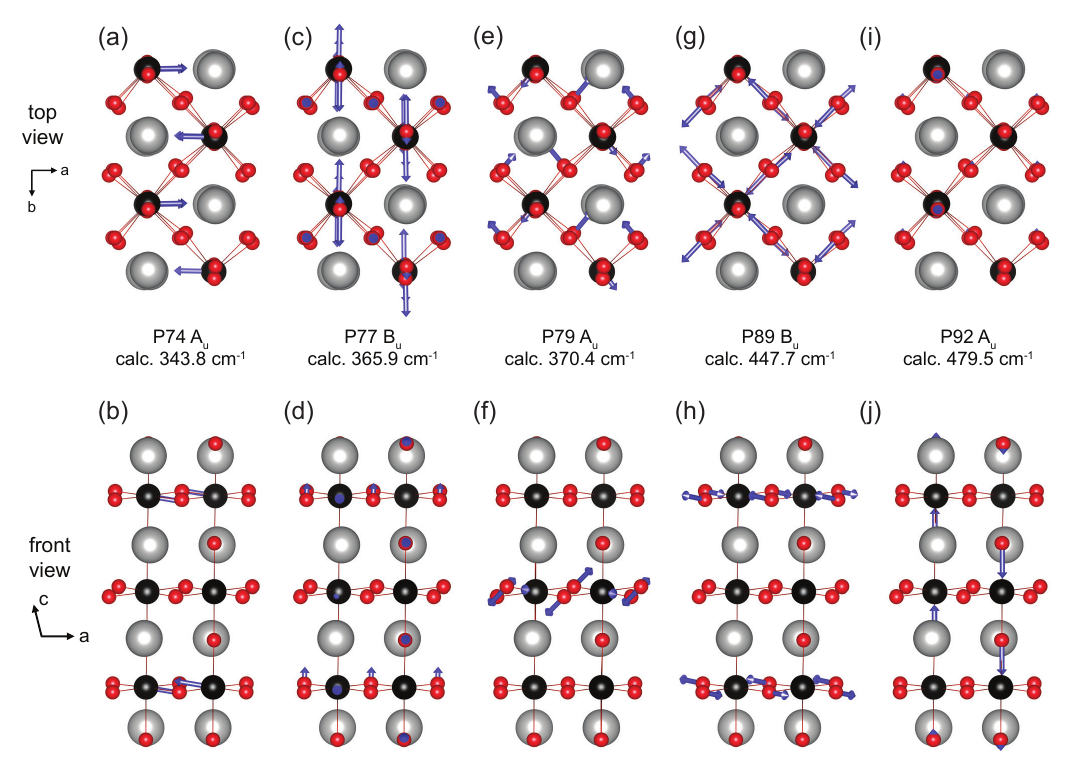}
    \caption{Calculated atomic displacement patterns for other infrared-active phonon modes in the frequency range between 300 and 500~cm$^{-1}$. Panels (a,b), (c,d), (e,f), (g,h), and (i,j) show the top and front views of the phonon eigenvectors for modes P74, P77, P79, P89, and P92, respectively. The corresponding calculated phonon frequencies are indicated for each mode. The mode labels (P\#) follow the ranking of the phonon frequencies among all 102 calculated zone-center phonon modes. For clarity, only displacements with large amplitude are shown for each mode. These representative eigenvectors illustrate that phonons in this frequency window primarily involve vibrations of the Ni-O bonds (including Ni-O stretching and bending character), consistent with the discussion in the main text.}
    \label{fig:S18_more_diplacement}
\end{figure}

\begin{figure}
    \centering
    \includegraphics[width=0.98\textwidth]{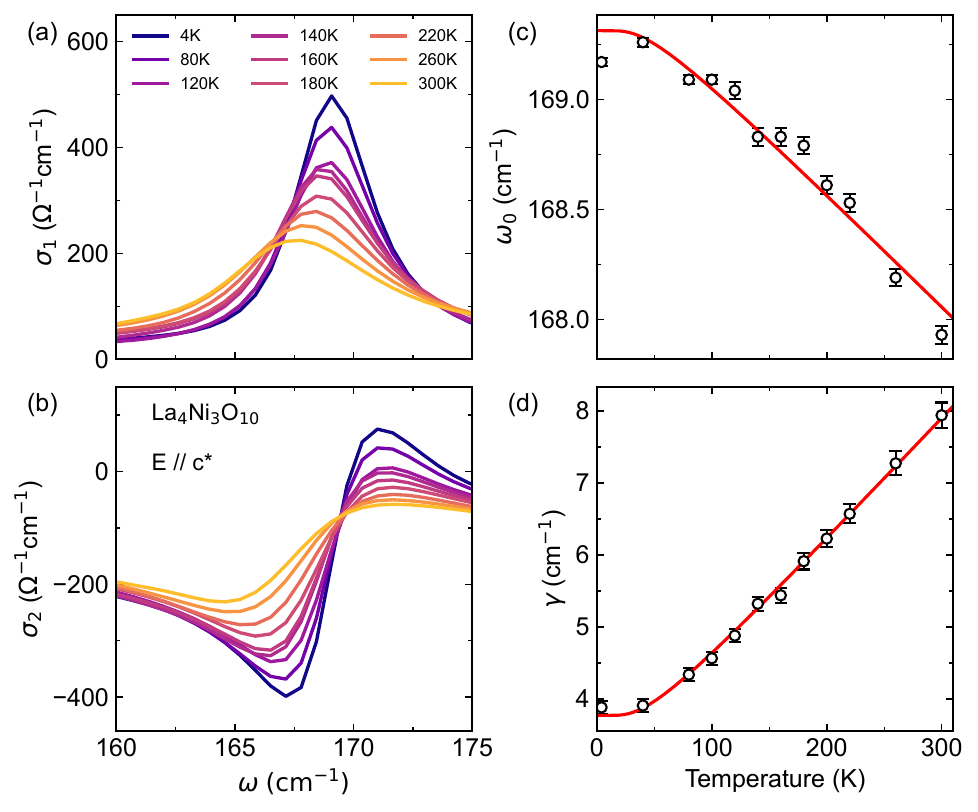}
    \caption{(a), (b) Real and imaginary out-of-plane optical conductivity of La$_4$Ni$_3$O$_{10}$, in a narrow frequency window around the 169~cm$^{-1}$ mode. (c), (d) Temperature dependence of the fitted phonon frequency $\omega_0$ and scattering rate $\gamma$ for the modes. In contrast to the pronounced anomalies of selected modes between 300 and 500~cm$^{-1}$, the phonon near 169~cm$^{-1}$ shows no discernible renormalization of its frequency or linewidth across the DW transition. Solid red curves show anharmonic phonon-decay fits over the full temperature range. The good agreement indicates that this mode is largely insensitive to the DW transition, underscoring the mode and energy selectivity of DW-induced phonon modulation.
}
    \label{fig:S17_P169_norenorm}
\end{figure}

The phonon anomalies induced by the DW transition are found primarily in the frequency range between 300 and 500~cm$^{-1}$. We therefore focus on four phonon modes in this window that exhibit the most pronounced deviations from conventional anharmonic behavior. Two modes, located near 371 cm$^{-1}$ (Fig. \ref{fig:S12_P375_split}) and 340 cm$^{-1}$ (Fig. \ref{fig:S13_P340_split}) in the high-temperature phase, split into distinct branches upon cooling to the DW state. In addition, two higher-frequency modes near 430 cm$^{-1}$ (Fig. \ref{fig:S14_P430_renorm}) and 475 cm$^{-1}$ (Fig. \ref{fig:S15_P475_renorm}) show strong renormalization without a resolvable splitting. For all four modes, the temperature dependence of the resonance frequency and linewidth is analyzed using the anharmonic decay model described above. By contrast, phonon modes outside this frequency window, such as the mode near 169~cm$^{-1}$ shown in Fig.~\ref{fig:S17_P169_norenorm}, exhibit temperature evolutions that remain well described by conventional anharmonic behavior over the full temperature range. This comparison highlights the non-uniform and energy-selective nature of the DW–phonon coupling, with the strongest deviations from anharmonicity occurring for modes in the frequency range between approximately 370 and 430 cm$^{-1}$. Notably, this range coincides with the typical energy scale of Ni–O bond–stretching vibrations, suggesting that lattice distortions involving Ni–O bond are particularly sensitive to the electronic modulation associated with the DW transition. To further elucidate the lattice degrees of freedom underlying the phonon anomalies, we examine the calculated displacement patterns of the four representative modes discussed above. As shown in Fig.~\ref{fig:S16_more_diplacement}, these modes are dominated by Ni-O bond bending, suggesting that Ni-O distortions are especially sensitive to the electronic reconstruction across the DW transition.
Consistent with this interpretation, additional DFT calculated eigenvectors for other modes in the same 300–500~cm$^{-1}$ window also exhibit substantial participation of the Ni-O bonds, even though these modes are not associated to experimentally observed phonons (Fig.~\ref{fig:S18_more_diplacement}). 

\clearpage
\bibliography{ref} % refs.bib